\documentclass[useAMS,usenatbib]{mn2e}
\usepackage{graphics}
\usepackage{graphicx}
\usepackage{txfonts}
\usepackage{color}

\title[The microturbulent precursor of relativistic shocks]{Particle
  transport and heating in the microturbulent precursor of
  relativistic shocks} \author[I.~Plotnikov, G.~Pelletier and
M.~Lemoine]{ Illya~Plotnikov,$^{1}$\thanks{e-mail:{\tt
      illya.plotnikov@obs.ujf-grenoble.fr}} Guy
  Pelletier,$^{1}$\thanks{e-mail:{\tt
      guy.pelletier@obs.ujf-grenoble.fr}} and
  Martin Lemoine,$^{2}$\thanks{e-mail:{\tt lemoine@iap.fr}}\\
  $^{1}$ UJF-Grenoble 1 / CNRS-INSU, Institut de Plan\'etologie et
  d'Astrophysique de Grenoble (IPAG) UMR 5274,
  Grenoble, F-38041, France.\\
  $^{2}$ Institut d'Astrophysique de Paris, CNRS, UPMC,
  98 bis boulevard Arago, F-75014 Paris, France\\
}

\date{Accepted ----. Received ----; in original form ----}

\volume{000}
\pagerange{000--000}
\pubyear{0000}

\begin{document}

\maketitle

\begin{abstract}
  Collisionless relativistic shocks have been the focus of intense
  theoretical and numerical investigations in recent years. The
  acceleration of particles, the generation of electromagnetic
  microturbulence and the building up of a shock front are three
  interrelated essential ingredients of a relativistic collisionless
  shock wave.  In this paper we investigate two issues of importance
  in this context: (1) the transport of suprathermal particles in the
  excited microturbulence upstream of the shock and its consequences
  regarding particle acceleration; (2) the preheating of incoming
  background electrons as they cross the shock precursor and
  experience relativistic oscillations in the microturbulent electric
  fields. We place emphasis on the importance of the motion of the
  electromagnetic disturbances relatively to the background plasma and
  to the shock front.  This investigation is carried out for the two
  major instabilities involved in the precursor of relativistic
  shocks, the filamentation instability and the oblique two stream
  instability. Finally, we use our results to discuss the maximal
  acceleration at the external shock of a gamma-ray burst; we find in
  particular a maximal synchrotron photon energy of the order of a few
  GeV.
\end{abstract}

\section{Introduction}
The microphysics of collisionless relativistic shocks has been
intensively investigated in recent years, through both numerical
simulations and theoretical investigations. As demonstrated in
particular by \cite{S08a,S08b}, the physics of these shock waves in
the unmagnetized limit involves the interplay of three phenomena: the
formation of the shock through the deceleration and reflection of
particles against a microturbulent magnetic barrier, the
self-generation of this microturbulence upstream of the shock by back
scattered particles and the development of Fermi type acceleration.

So far, particle acceleration has been observed in PIC simulations of
unmagnetized relativistic shocks
(\citealp{S08b,Kea09,Mea09,SS09,SS11a}), and indeed one must expect
the development of the Fermi process when the magnetization is very
weak, because microturbulence can then grow and provide the necessary
scattering (\citealp{LPR06,LP10}). At larger levels of magnetization
of the upstream flow, the shorter precusor scale may prevent the
development of microinstabilities, and in the absence of cross-field
scattering, Fermi powerlaws cannot develop (\citealp{Nea06,LPR06});
this picture has been validated in particular by the simulations of
\cite{SS11a}.

Nevertheless, the long timescales and high energies that are inferred
in powerful astrophysical sources remain well out of reach of these
state of the art numerical simulations. It is therefore important to
build on the basis of these numerical experiments a theoretical
understanding of the various processes at play in these shock waves.
In the present work, we are interested in the physics of the
microturbulence upstream of a relativistic weakly magnetized
shock. Two fast growing microinstabilities have received significant
attention regarding the development of microturbulence upstream of a
relativistic shock front: the filamentation (often termed Weibel) mode
(e.g. \citealp{ML99,WA04,LE06,AW07a,AW07b,B09,LP10,Rea11,LP11a,Sea11})
and the two stream instability (hereafter OTSI,
e.g. \citealp{BFP05,B09,LP10,LP11a,Sea11}). Both instabilities follow
from the interpenetration of the beam of back scattered particles and
the incoming background plasma in the shock precursor (as viewed from
the shock frame). One should nevertheless mention the possibility of a
Buneman instability, if the returning particles carry a net current,
which turns out to grow faster than the previous two, see
e.g. \cite{B09} and \cite{LP11a} for a discussion. However, the
Buneman instability saturates through the heating of the background
electrons, so that it presumably serves as an efficient source of
preheating.  In the present work, we are mostly interested in the
properties of particle transport (and energization) in the
microturbulence upstream of a relativistic shock front and we will
focus our discussion on the respective roles of the filamentation and
two stream modes.

In the downstream, the microturbulence appears isotropic, mostly
magnetic and static, see e.g. \cite{Cea08}. The physics of transport
of suprathermal particles in such a microturbulence, possibly
superimposed on a weak background magnetic field, has been discussed
in a previous paper (\citealp{PL11}). Upstream of the shock, this
microturbulence is strongly elongated in the direction of the shock
normal and in the background plasma rest frame, it carries both
electric and magnetic fields. That must affect the transport
properties of suprathermal particles in a non-trivial way and likely
contribute to the heating of background electrons. Furthermore, we
demonstrate in the present work that the filamentation modes have a
finite phase velocity in the background plasma rest frame, an issue
which to our knowledge has not been addressed before in the present
context. We find that this motion has important consequences regarding
both the transport of suprathermal particles, in particular the
acceleration timescale, and the preheating of the background
electrons, which turns out to be fast and efficient.

This paper is organized as follows. In Section~2, we discuss the
motion of the frame in which the Weibel filaments are static, and we
summarize previous findings on a similar issue for the OTSI mode. We
investigate the influence of the motion of the electromagnetic modes
on the reflection process at the shock front.  In Section~3 we study
the transport of suprathermal particles in both Weibel and OTSI
turbulence, on the basis of numerical simulations of test particle
propagation. We place emphasis on the issue of scattering in three
dimensions.  Section~4 discusses electron heating. We show that
the relativistic oscillation of the incoming background electrons in
the electric field of the microturbulence modes lead to efficient
preheating on a short timescale. In Section~5 we apply our results to
the concrete case of the external relativistic shock of a gamma-ray
burst.  We summarize our results in Section~6.

\section{Microturbulence in the precursor of collisionless weakly
  magnetized relativistic shocks} \label{sec2} 

We start with some definitions of key quantities. We note
$\mu\,\equiv\,m_e/m_p$ the electron to proton mass ratio, $\Gamma_{\rm
  s}$ the Lorentz factor of the shock, $n$ the density of the
background (upstream) plasma, $\rho\,=\,nm_pc^2$ the rest mass density,
$B_0$ the magnetic field of the background plasma; $\theta_B$ the
angle between the direction of $\mathbf{B_0}$ and the shock normal;
these quantities are defined in the upstream plasma rest frame. The
magnetization parameter is then defined as
\begin{equation}
\label{eq:mag}
\sigma\,\equiv\, \frac{B_{\rm t\mid f}^2}{4\pi \Gamma_{\rm s}^2 \rho c^2}\,
=\,  \frac{B_{\rm 0}^2}{4\pi \rho c^2} \sin^2 \theta_{\rm B}\ ,
\end{equation}
with $B_{\rm t\mid f}\, =\, \Gamma_s B_0 \sin \theta_B$ the transverse
component of the background magnetic field in the shock front rest
frame.  Numerical simulations provide two essential parameters for
astrophysical applications, namely the conversion factor $\xi_{\rm
  cr}$ of the incoming energy into cosmic rays (suprathermal
particles), and the conversion factor $\xi_{\rm B}$ into magnetic
energy:
\begin{equation}
  P_{\rm cr} \,\equiv\,  \xi_{\rm cr} \Gamma_{\rm s}^2 \rho c^2 \ ,\quad
  {\bar B_{\mid {\rm f}}^2 \over 4\pi} \, = \, \xi_{\rm B} \Gamma_{\rm
    s}^2\rho c^2 \ ,
\end{equation}
where the cosmic ray pressure $P_{\rm cr}$ and the level of magnetic
turbulence $\bar B_{\rm \mid f}$ are measured at the shock
front. These two crucial parameters $\xi_{\rm cr}$ and $\xi_{\rm B}$
are expected to be on the order of $1-10\,$\%
(\citealp{SS11a}). Actually the cosmic rays are considered, as shown
by numerical simulations and explained by theory, as the source of
magnetic, and more generally electro-magnetic, turbulence. In the
present context of a proton electron plasma of low magnetization, the
reference time scale is $\omega_{\rm pi}^{-1}$ and the spatial scale
of reference is the inertial scale of protons $\delta_{\rm i} \equiv
c/\omega_{\rm pi}$.


In this section we present the essential characteristics of the
filamentation instability and the oblique two stream instability,
which are the most important sources of turbulence in the precursor of
ultra-relativistic shocks. As briefly mentioned above, there are also
Buneman instabilities that are triggered by the compensation current
in the background plasma, which compensates the current carried by the
reflected particles either along the mean field (for a parallel shock)
or across the mean field. For instance, reflected particles of
opposite charge rotate in opposite direction in the transverse mean
field and thus produce a very intense diamagnetic current responsible
for a Buneman instability (\citealp{LP11a}), which grows
rapidly. Those current instabilities produce a turbulent heating of
the electrons up to some temperature that reduces the anisotropy of
the electron distribution function, up to the point where the
instability saturates. Therefore these current instabilities
participate in the process of preheating electrons, which then arrive
at the shock front with a relativistic temperature. In this paper, we
will not address the preheating effect due to Buneman instability and
focus on the Weibel and OTSI instabilities. In the picture that we
develop here, these instabilities indeed push the preheating to
higher temperatures, up to near equipartition, by the time the
electrons reach the shock front.

The first generation of reflected particles constitutes the main
content of suprathermal particles that penetrate the ambient plasma
with an energy density much larger than $\rho c^2$ as measured in the
frame of the ambient plasma.  Its interaction with the background
plasma can be described perturbatively as long as the proton beam of
(apparent) density $n_{\rm b}$ and Lorentz factor $\gamma_{\rm b}$ is
such that $\omega_{\rm pb} \ll \omega_{\rm p}$, which amounts to
$n_{\rm b} / \gamma_{\rm b} \ll n/ \mu$; now for a beam reflected by a
shock, $n_{\rm b} = \xi_{\rm cr} \Gamma_{\rm s}^2 n$ and $\gamma_{\rm
  b} \sim \Gamma_{\rm s}^2$; thus the weak interaction criterium
becomes $\xi_{\rm cr} \ll 1/\mu$, which is always realized.  The weak
interaction of the very energetic beam with the ambient (or upstream)
plasma renders the calculation of the instability convenient in the
frame of the background plasma.

\subsection{The frame of magnetic filaments}

Consider first the growth of magnetic perturbations due to the Weibel
instability triggered by a parallel beam of velocity $\upsilon_{\rm
  b}$ and Lorentz factor $\gamma_{\rm b} = (1-\upsilon_{\rm
  b}^2/c^2)^{-1/2}\sim\Gamma_{\rm s}^2$ (upstream frame), interacting
with a cold background plasma of density $n$ at rest; at a shock wave,
the beam of returning particles carries an energy density $\xi_{\rm
  cr} \gamma_{\rm b}^2 n m_p c^2$, so that the beam plasma frequency
$\omega_{\rm pb}\sim \xi_{\rm cr}^{1/2}\mu^{1/2}\omega_{\rm pe}$, with
$\omega_{\rm pe}$ the background electron plasma frequency (see
\citealp{LP10} for details). In the upstream rest frame, the e-folding
length scale of the instability is written
\begin{equation}
  \ell_{\rm g}  \,\equiv\, c/\Im \omega \sim \xi_{\rm cr}^{-1/2}\delta_{\rm i}\ ,
\end{equation}
the detailed growth rate (see appendix A) being
\begin{equation}
\label{eq:ginst}
\gamma_{\rm inst} \,\equiv\, \Im\omega \,= \,\sqrt{\xi_{\rm cr}}\omega_{\rm pi}  
\, \frac{\beta_{\rm b} k_{\perp} \delta_e}{(1+k_{\perp}^2\delta_e^2)^{1/2}} \ .
\end{equation} 
In the above expression and throughout this paper, $k_{\perp}$
represents the wavenumber component transverse to the shock normal,
i.e. tangential to the shock front, while the (longitudinal)
wavenumber component along the shock normal is written
$k_{\parallel}$.  For the filamentation mode, $k_{\perp}\gg
k_{\parallel}$. In general, one takes the limit
$k_{\parallel}\rightarrow 0$, which leads to an aperiodic mode, i.e.
$\Re \omega=0$.  However, for a small, but finite longitudinal wave
number $k_{\parallel}$, the Weibel modes have a non-vanishing real
frequency which indicates that these magnetic filaments propagate at
high velocity. Considering first the above case of a cold electron
background:
\begin{equation}
  \label{eq:or}
  \omega_r \,\equiv\,\Re\omega\,=\, 
k_{\parallel} \upsilon_{\rm b} \left( 1-\xi_{\rm cr} \mu\frac{\gamma_{\rm b}^{-2}
      +k_{\perp}^2\delta_e^2}{1+k_{\perp}^2\delta_e^2} \right) \ .
\end{equation}
We remark that this phase velocity is consistent with the result
obtained in the center of mass frame of the counter streaming
configuration when $\gamma_{\rm b} \simeq 1$; for $\mu = 1$ and
$\xi_{\rm cr} = 1/2$, we find a phase velocity $\upsilon_{\rm b}/2$ at
peak growth rate, which corresponds to a vanishing phase velocity in
the center of mass frame.

In this work we will often refer to the wave frame, which corresponds
to the frame moving at the phase velocity of the magnetic disturbance
$v_{\rm m}\,\equiv\,\beta_{\rm m}c$ along the shock normal direction:
\begin{equation}
  \beta_{\rm m}\,\equiv\,\frac{\Re \omega}{k_\parallel c}\ .
\end{equation}
By definition, the wave is static in this frame, since the mode
frequency in that frame $\omega' = \gamma_m(\omega - k_{\parallel}
v_m)=0$, with $\gamma_{\rm m}\,=\,\left(1-\beta_{\rm
    m}^2\right)^{-1/2}$. Stricly speaking this phase velocity is not
unique because of its dependence on $k_{\perp}$; however we consider
it for the most unstable mode, which introduces only a small dispersion
$\Delta \beta_m/\beta_m \sim \xi_{cr} \mu$. 

In the above cold background plasma limit, the Weibel filaments can be
considered as wave packets of transverse size $\ell_{\perp}\sim
\delta_e$ and extension in the normal direction $\ell_{\parallel} >
\delta_e$, but necessarily finite (at least because of the finite
growth length in a precursor of finite extent). Equation~\ref{eq:or}
indicates that these filaments move at a high speed relatively to the
ambient medium.  Indeed
\begin{equation}
\gamma_{\rm m} \,\simeq\, (\xi_{\rm cr} \mu)^{-1/2}\sim 140\ ,
\end{equation}
where the numerical value holds for an electron-proton plasma with
$\xi_{\rm cr}=0.1$. This is an important point, which has not been
taken into account in the literature, to our knowledge. The motion of
the wave packets indeed carries particular importance when determining
the condition for particle reflection, the level of electron
preheating and also for the analysis of the Fermi process, as
discussed in the following.

In the upstream rest frame, these waves are quasi-electromagnetic
waves with the electric field ${\bf E}$ mostly oriented along the beam
(shock normal), a dominant magnetic component perpendicular to both
${\bf k}_{\perp}$ and ${\bf E}$, and finally a small electrostatic
component of ${\bf E}$ along ${\bf k}_{\perp}$.  The magnetic
component remains the most intense component despite a phase velocity
close to c along the normal direction because $E/B \simeq (\xi_{cr}
+k_{\parallel}^2 \delta_i^2)^{1/2}/(k\delta_i)$. The fact that they
have a phase velocity close to $c$ implies that they suffer negligible
Landau damping, at least as long as the thermal velocity of the
electrons is small.

The filaments are static in the frame moving at $\beta_{\rm m}$ with
respect to the background plasma, however in their rest frame they are
no longer predominantly magnetic. They actually possess an
electrostatic field of almost the same intensity as the magnetic
field, as a standard Lorentz transform shows: $E'_{\perp} \simeq
\gamma_{\rm m} B\simeq B'$, since $B' = \gamma_{\rm m} B$, with the
prime denoting the quantity in the filament rest frame. This
electrostatic field turns out to be an important source of heating for
the background electrons, as discussed in Sec.~\ref{sec4}.

If the background electrons have been preheated to relativistic
temperatures, their electromagnetic response and therefore the
instability are modified.  The maximum growth rate remains unchanged,
$\Im\omega \sim \xi_{\rm cr}^{1/2}\omega_{\rm pi}$, but the spatial
scale at which maximum growth occurs is now larger because the
electron inertial scale is enlarged by a factor $\sqrt{\bar
  \gamma_e}$, with $\bar\gamma_e$ the mean electron thermal Lorentz
factor in the upstream rest frame $\bar \gamma_e =
1+3T_e/m_ec^2$. Detailed calculations are given in Appendix A; they
include in particular Landau damping on hot electrons, which
essentially reduces the growth rate at larger wavelengths.  This
latter spatial scale tends to $\delta_{\rm i}$ in the limit of
equipartition, meaning $\bar\gamma_e\rightarrow m_p/m_e$. In a
relativistically hot background, the phase velocity of the modes along
the beam direction is slowed down, but nonetheless remains
relativistic for reasonable values. The calculations presented in
Appendix~A indicate
\begin{equation}
\label{eq:gm}
\gamma_{\rm m} \,\simeq\, 
\left[\left(\bar \gamma_e \mu \xi_{\rm cr} \right)^{1/2} + 2 \xi_{\rm cr}\right]^{-1/2} \ .
\end{equation}
Equipartition corresponds to $\bar\gamma_e\mu=1$, therefore prior to
equipartition $\bar\gamma_e\mu<1$. For relativistically hot electrons
far from equipartion, $\gamma_{\rm m} \simeq (2\xi_{\rm cr})^{-1/2}$,
then $\gamma_{\rm m}\sim (\xi_{\rm cr} \mu \bar\gamma_e)^{-1/4}$ when
$\bar\gamma_e\mu>\xi_{\rm cr}$, and finally the lowest value of
$\gamma_{\rm m}\sim2$ is reached at equipartion. This value is
significantly smaller than that obtained in the cold electron
background, $\gamma_{\rm m}\sim (\xi_{\rm cr}\mu)^{-1/2}$, but it
nevertheless plays an important role in what follows.

Note that the filamentation instability in the usptream of a
relativistic shock is quenched when $\Gamma_{\rm s} < (\xi_{\rm cr}
\mu)^{-1/2}$ as long as the electrons are kept cold due to the finite
angular dispersion of the beam (\citealp{Rea11}, \citealp{LP11a}).  In
a hot background however, the filamentation instability remains
strong, even when the beam angular dispersion is taken into account.

The above discussion shows that the velocity of the filaments depends
on the degree of preheating of the background electrons, hence on the
distance to the shock front, since electrons are cold at the tip of
the precursor and hot at the shock transition, see \cite{SS11a}.
Three cases deserve attention: $\gamma_{\rm m} \sim \Gamma_{\rm s}$
where the filaments move more or less at the same speed as the shock
front, $\gamma_{\rm m} < \Gamma_{\rm s}$ where the shock front catches
up the filaments and the case where the filaments can even run faster
than the shock for $\Gamma_{\rm s} < \gamma_{\rm m}$. In this latter
case the generation of filaments might lead to shock reformation, but
this issue is not discussed here.

The relative velocity of the filaments with respect of the shock front
can be written
\begin{equation}
\label{ }
\beta_{\rm m \mid f} \,\simeq\,  \frac{\gamma_{\rm m}^2-\Gamma_{\rm s}^2}{\gamma_{\rm m}^2 + \Gamma_{\rm s}^2} \ ,
\end{equation}
and the associated Lorentz factor
\begin{equation}
\label{eq:gmf}
\Gamma_{\rm m\mid f} \,\simeq\,{1\over 2}\left({\gamma_{\rm m} \over \Gamma_{\rm s}}+{\Gamma_{\rm s} \over \gamma_{\rm m}}\right) \ .
\end{equation}

The discussion about the motion of magnetic disturbances is also
important for their transmission to the downstream flow. Indeed the
relative motion of the upstream flow with respect to the downstream
one is characterized by a relative Lorentz factor $\Gamma_{\rm r} =
\Gamma_{\rm s}/{\sqrt 2}$. The Lorentz transform of the frequency and
normal wavenumber of filaments to the downstream frame leads to
\begin{eqnarray}
  \omega_{\vert\rm d} & = & \Gamma_{\rm r}(\beta_{\rm m}-\beta_{\rm
    r})k_\parallel c \simeq {1\over 2} \Gamma_{\rm r}k_\parallel c 
  \left({ 1\over \Gamma_{\rm r}^2} - {1\over \gamma_{\rm m}^2} \right ) \nonumber\\
  k_{\parallel\vert\rm d} & = & \Gamma_{\rm r}(1-\beta_{\rm r} \beta_{\rm m}) k_\parallel \simeq {1\over 2} \Gamma_{\rm r} k_\parallel
  \left({ 1\over \Gamma_{\rm r}^2} + {1\over \gamma_{\rm m}^2} \right )
\label{eq:transmission}
\end{eqnarray}
Thus, when $\gamma_{\rm m} \ll \Gamma_{\rm s}$, the modes are
perceived as electromagnetic vacuum waves in the downstream frame,
propagating backwards ($\omega_{\vert\rm d} \simeq
-k_{\parallel\vert\rm d}c$). However, one should stress
here that the motion of the filament is calculated at the linear
level, while the transmission of modes downstream proceeds in the
highly nonlinear regime. Therefore it is not clear at present whether
the above Eq.~\ref{eq:transmission} applies to the transition; it
should be taken with caution.

Finally, in the case of an ultra-relativistic shock in an electron
positron plasma, $\gamma_{\rm m}$ is at most a few and the shock front
catches up the magnetic disturbances ($\Gamma_{\rm s} \gg \gamma_{\rm
  m}$). The behavior is the same in a proton-electron plasma at near
equipartition.

\subsection{The role of OTSI turbulence}
Although the Weibel instability can produce all the effects expected
to occur at a relativistic shock, the OTSI appears unavoidable, and it
grows a little bit faster than the Weibel instability at least in the
cold background limit. The OTSI is a resonant instability of electron
plasma waves ($\omega \simeq \omega_{{\rm p}e}$) with a sharp
selection of the wave-vector component along the beam:
$k_\parallel=\omega_{{\rm p}e}/c$, which grows much faster than the
usual two stream instability when the transverse component of the
wave-vector is also of order $\omega_{{\rm p}e}/c$
(e.g. \citealp{F70}, \citealp{BFP05}). These modes are mostly
electrostatic in the background plasma frame; their frequency is
slightly shifted off resonance according to: $\Re\omega= \omega_{{\rm
    p}e}(1-{\vert \delta \vert / 2})$ and their growth rate $\Im\omega
= {\sqrt{3}\, 2^{-4/3}} \vert \delta \vert \omega_{{\rm p}e}$ with
$\vert \delta \vert = \xi_{\rm cr}^{1/3}\mu^{1/3}$
(e.g. \citealp{LP10,LP11a}).  The ratio of the electromagnetic
component over the electrostatic one is of order $\vert \delta \vert$.

As discussed in \cite{LP11a}, \cite{Sea11}, the two stream instability
becomes inhibited once the electrons are heated to ultra-relativistic
temperatures and the Weibel instability becomes the dominant mode.

In order to study the dynamics of particles in the modes, it is
interesting to move to the wave frame in which the particles
experience a static electric field and a static magnetic field. The
wave frame velocity with respect to the background plasma $\beta_{\rm
  m} = 1-\vert \delta \vert/ 2$ and the corresponding Lorentz factor
\begin{equation}
  \gamma_{\rm m} \,\simeq\,  \vert \delta \vert^{-1/2} \simeq \xi_{\rm cr}^{-1/6}
  \mu^{-1/6}\sim 5
\end{equation} 
with respect of the background plasma. These mildly relativistic
wave packets are thus rapidly overtaken by the shock front. 

As in the case of Weibel modes, the dynamics in OTSI modes is also
governed by a couple of transverse fields of similar amplitude when
one shifts to the rest frame in which these modes are static. A
crucial difference however is that the OTSI mode appears as a high
frequency quasi vacuum monochromatic wave in the front frame,
revealing a clear periodic pattern in the direction of the shock
normal.

\subsection{Suprathermal particles, the background plasma and the
  shock}\label{sec:stb}
In Sections~\ref{sec3} and \ref{sec4} that follow, we address
respectively the issues of the scattering of suprathermal particles
and the heating process of background particles in the motional
microturbulence upstream of the shock. These actually represent two
different facets of a similar problem, namely particle transport in a
time varying microturbulence. However, in the test particle picture
that we adopt in the following, these two populations, the
suprathermal particles and the background electrons, differ one from
the other by their wiggler parameter
\begin{equation}
a \,\equiv\, {e\bar E' \ell_{\perp}'\over m_ec^2} \ ,
\end{equation}
as expressed in terms of the microturbulent electric field $\bar E'$
and transverse scale $\ell_{\perp}'$ in the wave frame. As discussed in the
following, in the wave frame $\bar E'$ and $\bar B'$ are of the same
order, so that there is no ambiguity in the definition of $a$.

Cold background electrons have a Lorentz factor $\gamma'\sim \gamma_{\rm m}$
in the wave frame, so that 
\begin{equation}
\frac{a}{\gamma'}\,\simeq\,
\xi_{\rm B}^{1/2}\frac{m_p}{m_e}\frac{\ell_{\perp}}{\delta_{\rm i}}\,\gg\,1\ ,
\end{equation}
meaning that these electrons experience relativistic oscillations on a
coherence length scale in the wave frame. Of course, as the electrons
near equipartition, the ratio $a/\gamma'$ becomes closer to unity.

In sharp contrast, the same wiggler parameter for suprathermal
particles $a/\gamma'\ll 1 $. To see this, consider a suprathermal
electron, with Lorentz factor $\gamma$ in the upstream rest frame,
becoming $\gamma'\simeq\gamma/\gamma_{\rm m}$ in the mode rest frame
(assuming $\gamma\gg\gamma_{\rm m}$). The minimal Lorentz factor of
suprathermal electrons is $\gamma_{\rm min}\sim \Gamma_{\rm
  s}^2m_p/m_e$ in the upstream frame, hence one can write
\begin{equation}
  \frac{a}{\gamma'}\,
  \simeq\,\xi_{\rm B}^{1/2}\frac{\gamma_{\rm m}^2}{\Gamma_{\rm s}^2}\frac{\gamma_{\rm
      min}}{\gamma} \frac{\ell_{\perp}}{\delta_{\rm i}}\  ,
\end{equation}
which is indeed expected to be much smaller than unity: recall that
$\ell_{\perp}$ is expected of the order of $\delta_e$ if the
background electrons are cold (in which case $\gamma_{\rm m}$ can be
large if the Weibel instability is not quenched by the angular
dispersion of the beam), but of order $\delta_{\rm i}$ if the
background electrons reach equipartition with the ions, in which case
$\gamma_{\rm m}$ becomes of order of a few for the Weibel modes.  Thus
$a/\gamma'\ll 1 $ for suprathermal electrons, while $a/\gamma'\gg 1$
for the background electrons, typically.

For similar reasons, the microturbulence cannot trap the
suprathermal protons and thereby saturate the ion filamentation
instability that these particles seed. This would require that the
level of microturbulence is such that the time scale of non-linear
oscillation $\tau_{\rm nl}$ in the filament becomes comparable to the
growth timescale $\Im\omega\sim\omega_{\rm pb}^{-1}$. This oscillation
timescale can be expressed as: $\tau_{\rm nl}\,=\, \ell_{\perp}'
\left(\gamma'/a\right)^{1/2}/c$ and $\omega_{\rm pb} = \xi_{\rm
  cr}^{1/2} \omega_{\rm pi}$, so that one would need $a/\gamma'
\,\simeq\, \xi_{\rm cr}^{-1}(\ell_\perp/\delta_i)^2$ for suprathermal
particles, which cannot be satisfied. A similar conclusion can be
drawn when one considers the resonance broadening effect, which stems
from the fact that particle scattering or diffusion broadens the
resonance of modes with the beam responsible for the
instability. 

Presumably, the Weibel instability in the shock precursor does not
actually saturate, but it stops growing once the incoming (background)
plasma ions are turned around by the microturbulent field, as viewed
in the shock front rest frame. At this point, the shock transition
actually takes place. This notably implies that the back scattered
particles are roughly isotropic at this location (again, as viewed in
the shock front frame). This and the near isotropy of the incoming
ions, as they are turned around, then imply the end of growth of the
Weibel instability. Interestingly, this argument leads to a level of
magnetic turbulence which agrees well with current simulations.

This can be seen as follows, in the upstream rest frame in which the
filamentation modes are mostly static. In a first approximation, the
transverse magnetic field can be described as coherent in the shock
normal direction on a growth length scale $c/\Im\omega$. An incoming
proton is turned around on this length scale provided
\begin{equation}
\frac{e\bar B}{m_pc} \approx \Im\omega\ ,
\end{equation}
and $\Im\omega \,=\,\omega_{\rm pb}$ for the filamentation instability
implies a level
\begin{equation}
\frac{\bar B^2}{4\pi} \,\approx\,  \xi_{cr} nm_pc^2 \ ,
\end{equation}
meaning $\xi_B\,\approx\,\xi_{\rm cr}$ at the shock front.  The above
agrees rather satisfactorily with the PIC simulations of
\cite{SS11a}. Of course, the actual time to reach this amplitude,
starting from some background fluctuation value far upstream, is of
order $\Im\omega$ times an e-folding factor of order of a few to ten.

\section{Transport of suprathermal particles in the wave frame}\label{sec3}
The properties of particle transport in microturbulence has been
already studied downstream of a relativistic shock
(e.g. \citealp{Cea08,PL11}).  There, the transport coefficients are
found to depend essentially on a scattering frequency, which scales as
the square of the particle energy, corresponding to small pitch-angle
scattering:
\begin{equation}
  \nu_{\rm s\vert\rm d} = {2\over 3}{c \over \ell_{\rm c\vert d}} \left( \frac{e\bar
      B_{\vert\rm d}}{\epsilon_{\rm d}} \right )^2 \ ,
\end{equation}
with $\ell_{\rm c\vert d}$ the coherence scale, $\bar B_{\vert\rm d}$
the total magnetic field and $\epsilon_{\rm d}$ the particle energy,
in the downstream frame.  In the absence of a mean field, spatial
diffusion is isotropic and is simply described by the standard
diffusion coefficient
\begin{equation}
D \,\simeq\, c\ell_{\rm c\vert d} \left(\frac{\epsilon_{\vert\rm d}}{e \bar
    B_{\vert\rm d} \ell_{\rm c\vert d}} \right )^2 \ .
\end{equation}
However spatial diffusion is anisotropic in the presence of a mean
field; then the transverse diffusion coefficient tends toward a
constant value as the particle energy becomes large (\citealp{PL11}).

Upstream, the situation is rather different, notably because of the
anisotropy of the microturbulence.  

In the upstream frame, the wavenumbers of Weibel modes obey
$k_{\perp}/k_{\parallel}\gg 1$ and $\Re\omega\ll k_{\perp}$, so that
$k_{\perp}'/k_{\parallel}'\gg1 $ in the wave frame as well. Regarding
OTSI modes, one finds $k_{\perp}\sim k_{\parallel}$ in the upstream
frame, but Lorentz boosting to the wave frame gives predominance to
the transverse wavenumbers. If, in a first approximation, the spatial
dependence along the shock normal is disregarded, the normal component
of the generalized momentum becomes a constant of motion in the wave
frame: $p_{\parallel} +eA_\parallel(x,y) = C$ ($C$ constant,
$A_\parallel$ parallel component of the electromagnetic vector
potential). In this frame the total energy is also a constant of
motion due to time translation invariance. Thus the momentum of the
particle is confined in a subset of the energy surface determined by
$\Delta p_{\parallel} = e \Delta A_\parallel$, where $\Delta
A_\parallel$ is the rms variation of the normal vector potential (wave
frame). In short, the assumption of translational invariance along the
Weibel filaments leads to an inhibition of momentum diffusion, see
also \cite{JJB98} for similar issues. This is a crucial point which
directly impacts the efficiency of acceleration, which requires
transverse scattering in the absence of a mean field. As we discuss
further below, the transverse momentum is subject to large angular
variations in the transverse plane, which leads to spatial diffusion.
In order to obtain pitch angle diffusion, it is thus necessary to
consider the full 3D dependence of the magnetic fluctuations.

For both Weibel and OTSI modes, the analysis of particle dynamics is
more suitable in the wave frame, because this is the frame in which
the transport coefficients can be properly defined, where the
distribution function tends to become more or less isotropized, and
fundamentally, this is the proper frame of scattering centers involved
in the Fermi process. Henceforth, all quantities are therefore
evaluated in the frame of magnetic disturbances, unless otherwise
stated.

The electromagnetic components in the wave frame are derived from
thoses calculated upstream at the linear level by the Lorentz
transforms:
 \begin{eqnarray}
   \bmath{E}_{\parallel}' & = & \bmath{E}_{\parallel}\nonumber\\
   \bmath{E}_{\perp}' & = & \gamma_{\rm m} (\bmath{E}_{\perp} + \bmath{\beta}_{\rm m} \times \bmath{B}_{\perp})\nonumber \\  
   \bmath{B}_{\parallel}' & = & \bmath{B}_{\parallel} \nonumber\\
   \bmath{B}_{\perp}' & = & \gamma_{\rm m} (\bmath{B}_{\perp} -
   \bmath{\beta}_{\rm m} \times \bmath{E}_{\perp})\ .
\end{eqnarray}
In the case of Weibel modes, $\vert E \vert \ll \vert B \vert$, and
for $\gamma_{\rm m}$ large enough, $\bmath B'_{\perp} \simeq
\gamma_{\rm m} \bmath B_{\perp}$ and $\bmath E'_{\perp} \simeq
\gamma_{\rm m} \bmath \beta_{\rm m} \times \bmath B_{\perp} \simeq
\bmath \beta_{\rm m} \times \bmath B'_{\perp}$.
 
In the case of OTSI modes, in the upstream frame the modes are almost
electrostatic $\vert \bmath B \vert \ll \vert \bmath E \vert $, and
$E_{\perp}\sim E_{\parallel}$ (oblique modes). The system is quite
similar to the system derived for Weibel modes in their proper frame,
since $\bmath E'_{\perp} \simeq \gamma_{\rm m} \bmath E_{\perp}$ and
$\bmath B'_{\perp} \simeq -\gamma_{\rm m} \bmath \beta_{\rm m} \times
\bmath E_{\perp}$, which leads to $\bmath B'_{\perp} \simeq -\bmath
\beta_{\rm m} \times \bmath E'_{\perp}$. 

Below, we analyze the particle dynamics first in a 2D approximation,
meaning $E_{\parallel}'\rightarrow 0$, $B_{\parallel}'\rightarrow 0$,
and then in the complete 3D configuration
($E_{\parallel}'=E_{\perp}'/\gamma_{\rm m}$). When going to the wave
frame, the perpendicular coherence length remains unchanged,
$\ell_{\perp}'=\ell_{\perp}$, while $\ell_{\parallel}'=\gamma_{\rm
  m}\ell_{\parallel}$.

\subsection{Transport in 2D approximation}
Defining as $\mathbf{z}$ the direction of the shock normal, the system
can be written:
\begin{eqnarray}
  \frac{{\rm d}p_x}{{\rm d} t} & = & q E_x' (1+\beta_{\rm m} \beta_z)\\
  \frac{{\rm d}p_y}{{\rm d} t} & = & q E_y' (1+\beta_{\rm m} \beta_z)\\
  \frac{{\rm d}p_z}{{\rm d} t} & = & - q \beta_{\rm m} (\beta_x E_x' + \beta_y E_y')\label{eq:sys}
\end{eqnarray}  
where the relation $\bmath B_{\perp}' = -\bmath \beta_{\rm m} \times
\bmath E'_{\perp}$ has been inserted, and the fields $E_{\parallel}'$
and $B_{\parallel}'$ have been discarded in a first approximation.
Because $\omega' = 0$ or $\omega = k_z v_m$, $k'_z = \gamma_{\rm
  m}(k_z - \beta_{\rm m} \omega/c) = k_z/\gamma_{\rm m}$, and thus for
large $\gamma_{\rm m}$ the z dependence of the field can be neglected
in a first approximation. The system is the same for both Weibel and
OTSI modes.
   
As discussed above, there are two invariants: the total particle
energy $H= \epsilon(p) +q\Phi(x,y)$ written in terms of a Hamilton
function with electromagnetic potential $\Phi$, and the generalized
momentum component along $z$, $\pi_z = p_{\parallel} + qA_z(x,y)/c$.
These two potentials are related to one another: $A_z(x,y)
=\Phi(x,y)$. Since the potential has a zero average and a finite rms
value $\Delta \Phi$, the particle proper energy $\epsilon$ and its
z-momentum $p_{\parallel}$ have well defined rms variations under the
ergodic assumption: $\Delta \epsilon = \Delta p_{\parallel} c =
e\Delta \phi$. These relations determine confinement regions in phase
space and forbid some diffusion processes.  The norm of the transverse
momentum also has bounded variations. Indeed, considering the
variation from initial values, $\delta \epsilon =
\epsilon-\epsilon_0$, $\delta p_{\parallel} =
p_{\parallel}-p_{\parallel,0}$, $\delta \bmath p_{\perp} = \bmath
p_{\perp}- \bmath p_{\perp,0}$, one exactly finds
\begin{equation}
  2(\epsilon_0-p_{\parallel,0}c) \delta \epsilon = c^2 (\delta \bmath p_{\perp})^2 \ .
\end{equation}
Now because $p_{\perp,0}$ is small, this constraint allows large
variations of the polar angle of $\bmath p_{\perp}$. Thus the
variations of the energy and momenta are bounded, except for the angle
of the transverse momentum that can vary randomly over the interval
$(0, \, 2\pi)$; those erratic variations of the angle can occur with
the $\beta_z$ contribution to the transverse equations of motions that
opens phase space with another degree of freedom.

It proves convenient to define the reference energy
\begin{equation}
\epsilon_\star \equiv e\bar E' \ell_{\perp} \equiv a m_e c^2\ .
\end{equation}
As discussed before, $a/\gamma'\ll1 $ for suprathermal particles,
meaning that $\epsilon_\star\ll \epsilon$. Such particles are strongly
beamed forward along the shock normal in the upstream frame; in the
wave frame, $p_{\perp,0}/p_{\parallel,0}\lesssim \gamma_{\rm m\vert
  f}/\Gamma_{\rm s}$ implies $p_{\perp,0}\,\ll\,p_{\parallel,0}$ for
typical values of $\gamma_{\rm m\vert f}$, hence $\epsilon_0\,\simeq\,
p_{\parallel,0}c$ initially. The transport of suprathermal particles
can then be described as the random walk of a \emph{non-relativistic}
particle in the transverse plane, coupled to ballistic motion along
the longitudinal direction. In the transverse plane, the particle is
described as non-relativistic because its transverse velocity
$\upsilon_{\perp}\simeq p_{\perp}/p_{\parallel}\,\ll\,1$ and, given
that $\epsilon_\star\ll p_{\parallel,0}c$, the particle cannot
exchange a large fraction of its parallel momentum with transverse
momentum due to the invariance properties in this 2D approximation.

The transverse motion, devoid of linear resonance but governed by a
continuum of Fourier modes, is thus characterized by a single
nonlinear time 
\begin{equation}
  t_{\rm nl} \equiv {\ell_{\perp}\over c}\left({2p_{\parallel,0}c \over
    \epsilon_\star}\right)^{1/2}\ .
\end{equation}
This is the time needed to cross a coherence cell for a particle that
gets accelerated transversely in the transverse electric field. Over a
coherence length, the electric field can be considered as constant and
the particle receives a transverse kick $c \Delta p_{\perp} \sim
\epsilon_\star$. 

If the initial transverse momentum $p_{\perp,0}\ll \epsilon_\star/c$, the
spatial transverse motion can be approximated by
\begin{equation}
\label{eq:dxt}
\Delta \bmath x_{\perp} \simeq {1\over 2} {e\bmath E_{\perp} \over p_{\parallel,0}} ct^2 + \bmath \upsilon_{\rm \perp,0} t \ ,
\end{equation}
since the kick remains much smaller than $(\epsilon_\star
p_{\parallel,0})^{1/2}$; $\bmath \upsilon_{\perp,0}\equiv \bmath
p_{\perp,0}/p_{\parallel,0}$. The nonlinear time so defined is the
time beyond which the nonlinear dynamics de-correlates the
trajectories.  It can also be considered as the time step for a random
deflection of angle $\theta_i$ in the transverse plane since each
crossing of a coherence cell in the transverse direction is associated
to a large variation of $\theta_i$ when $p_{\perp,0}\ll
\epsilon_\star/c$.  Thus after $n$ steps of size $\ell_{\perp}$, the
trajectory has diffused such that
\begin{equation}
\label{ }
\langle\Delta \bmath x_{\perp}^2\rangle \simeq {1\over 2} \ell_{\perp}^2 n \ ,
\end{equation}
with $n \simeq \Delta t/t_{\rm nl}$, which leads to a transverse spatial
diffusion coefficient
\begin{equation}
\label{eq:dt}
D_{\perp} \simeq {1\over 4} {\ell_{\perp}^2 \over t_{\rm nl}} \propto p_{\parallel,0}^{-1/2} \ .
\end{equation}

If $p_{\perp,0}c \gg \epsilon_\star$, the transverse velocity
undergoes small variation of its modulus in the crossing of a
coherence cell, but it can undergo significant angle variations. This
transverse quasi scattering can be analyzed in two regimes: (a) when
$p_{\perp,0}c \ll (\epsilon_\star p_{\parallel,0})^{1/2}$ and (b) when
$p_{\perp,0}c \gg (\epsilon_\star p_{\parallel,0})^{1/2}$ but still
$p_{\perp,0}\ll p_{\parallel,0} $. In the former limit, the particle
crosses a transverse coherence cell in a nonlinear time scale as
previously, whereas in the latter case the crossing occurs on a linear
timescale $\ell_{\perp}/\vert \upsilon_{\perp} \vert$. Let us estimate
the transverse diffusion coefficient for both cases. In case (a), when
crossing a coherence cell, the particle undergoes a small transverse
deflection of angle $\theta_i \sim {\epsilon_\star/p_{\perp}c}$. The
scattering time is therefore longer than $t_{\rm nl}$ with $t_{\rm
  scatt} \sim \theta_i^{-2} t_{\rm nl}$ and diffusion in the
transverse plane during $\Delta t > t_{\rm scatt}$ is such that
\begin{equation}
\label{eq:Dperp_OTSI_2 }
\langle\Delta \bmath x_{\perp}^2\rangle \sim {\ell_{\perp}^2 \over 2}
\left( {t_{\rm scatt} \over t_{\rm nl}} \right )^2 {\Delta t \over t_{\rm scatt}} \ .
\end{equation}
This indicates that, in this regime of small deflection, a particle travels over a much 
larger distance than $\ell_{\perp}$ during a scattering time, namely $(t_{\rm scatt}/t_{\rm nl}) \ell_{\perp}$. 
The transverse diffusion coefficient is thus
\begin{equation}
\label{eq:Dperp_OTSI_3 }
D_{\perp} \sim {1\over 4}{\ell_{\perp}^2 \over t_{\rm nl}} \left( {p_{\perp} c \over \epsilon_\star}\right)^2 \ .
\end{equation}
In case (b), the small deflection that occurs during the linear
correlation time leads to a scattering time $t_{\rm scatt} \sim
\theta_i^{-2} \ell_{\perp}/\upsilon_{\perp}$.  And the transverse diffusion coefficient
becomes:
\begin{equation}
\label{ }
D_{\perp} \sim {1\over 4} \upsilon_{\perp}\ell_{\perp} \left( {p_{\perp} c \over \epsilon_\star} \right)^2 \ .
\end{equation}

\subsection{Numerical results in 2D-approximation}

The system of equations of motions in the fields is solved by the
Bulirsch-Stoer algorithm (\citealp{Press86}), together with statistics
over a sample of random phases and polarization directions of plane
waves. The field $\bmath{E}_{\perp}'$ is decomposed in plane waves
with the constraint $\bmath{k'} \parallel \bmath{E}_{\perp}'$, which
implies that $\bmath{k'}$ is in the plane $(x,y)$, choosing as before
the normal direction along the $z-$direction. In this subsection we
set $k_z' =0$ and $E_z' = 0$ in the wave frame.  Several hundreds of
modes are used. The system converges quite rapidly with $\sim 10^3$
particles. For simplicity we investigate only the case were the
initial particle momentum is oriented along the parallel direction,
$\bmath{p}_0=p_{\parallel,0}\bmath{e}_{\rm z}$.

All simulations use the following units: the spatial length unit is $m_e
c^2 / (e E'_{\rm rms})$ ($E'_{\rm rms}$ is the root mean square of the
electric field strength), the time unit is $m_e c / (e E'_{\rm rms})$,
and momenta are expressed in units of $m_ec$.  The coherence length
$\ell_{\perp}=1$ in these units, so that $\epsilon_\star=m_ec^2$.

\begin{figure}
\begin{center}
 \includegraphics[width=0.45\textwidth]{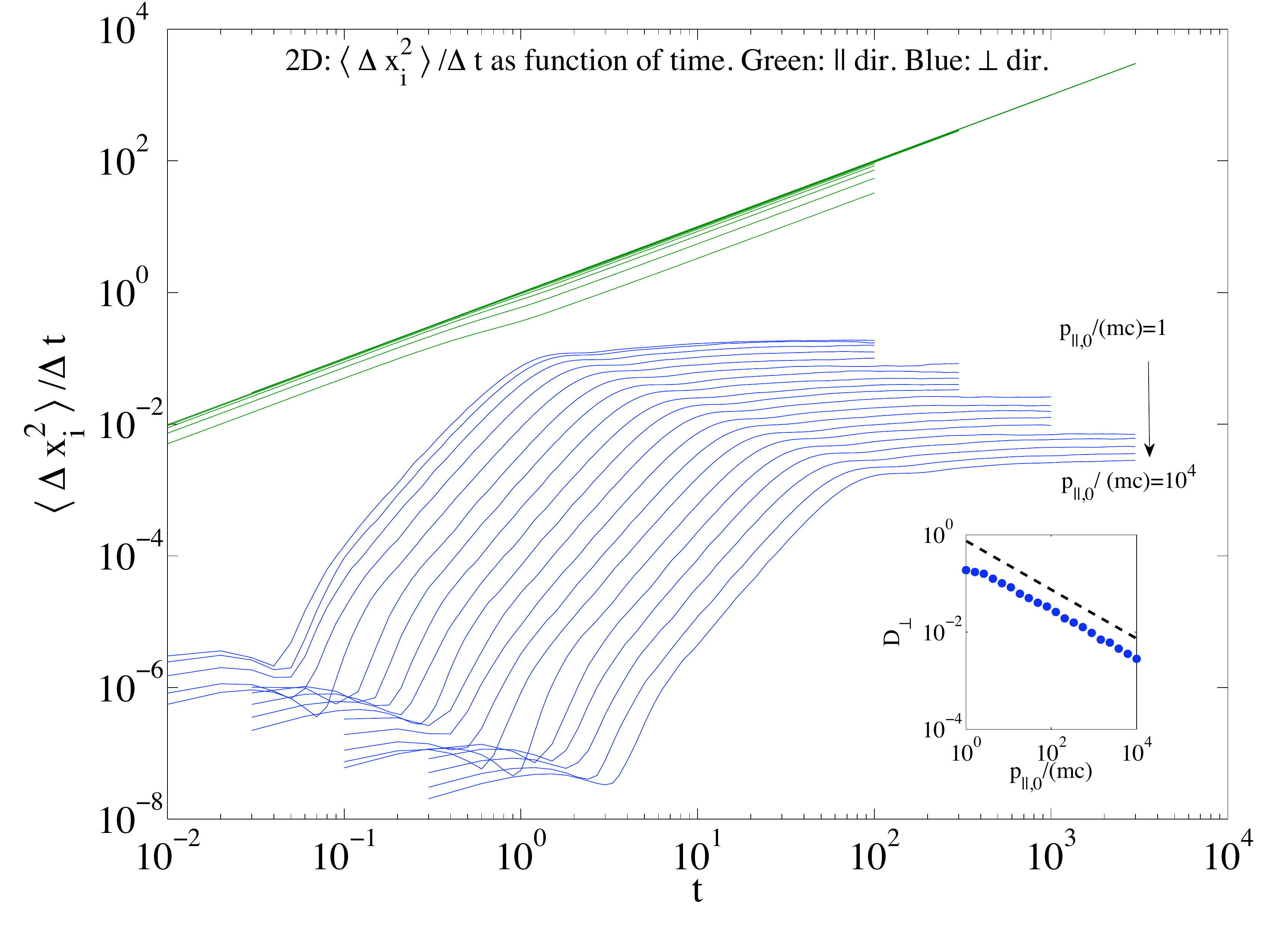}
 \caption{Time evolution of spatial diffusion coefficients for
   different $p_{\parallel,0}/(m_ec)$ in 2D fields geometry
   approximation. Green curves: parallel direction; Blue curves :
   transverse direction [$(x,y)$ plane]. Displacements in the parallel
   direction are clearly ballistic: $\langle\Delta x_{\parallel}^2
   \rangle =\upsilon_{\parallel,0}^2 t^2$. In the transverse plane, at least
   two regimes are identified: when $t<t_{\rm nl}$, $\langle\Delta
   x_{\perp}^2\rangle \propto t^4$ as expected from particle
   acceleration by nearly constant electric field over its coherence
   cell; and when $t>t_{\rm nl}$, $\langle\Delta x_{\perp}^2 \rangle /
   \Delta t$ reaches a plateau (i.e. diffusion) and its value has a
   power-law dependence on $p_{\parallel,0}$ with a slope of $-1/2$
   (see subpanel). Subpanel: Transverse diffusion coefficient
   (plateau) dependence on initial particle momentum
   $p_{\parallel,0}$. Dashed line follows the power-law slope $-1/2$.}
  \label{2D_1:fig1}
 \end{center} 
\end{figure}

\begin{figure}
\begin{center}
 \includegraphics[width=0.45\textwidth]{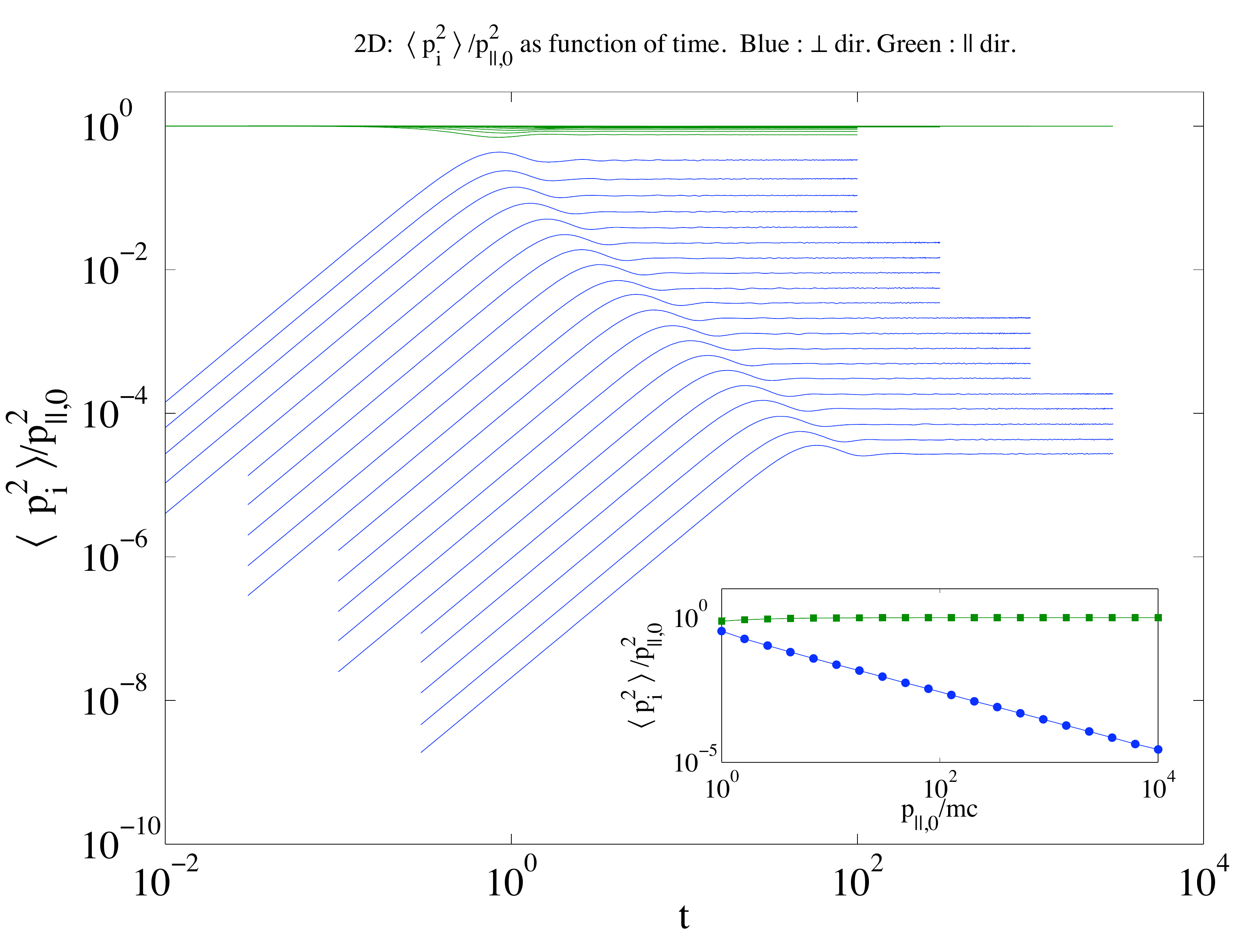}
 \caption{Time evolution of $\langle p_i^2 \rangle /p_{\parallel,0}^2$ for different values
   of $p_{\parallel,0}/(m_ec)$, in 2D approximation.  Green curves: parallel
   (z) direction; Blue curves : transverse direction. Subpanel:
   Asymptotic values of $\langle p_i^2 \rangle /p_{\parallel,0}^2$ as function of
   $p_{\parallel,0}$. $\langle p_{\parallel}^2 \rangle /p_{\parallel,0}^2$ is slightly inferior to 1
   independently from $p_{\parallel,0}$ (green squares). In transverse
   direction (blue circles) $\langle p_{\perp}^2 \rangle /p_{\parallel,0}^2 \propto
   p_{\parallel,0}^{-1}$.}
  \label{2D_2:fig2}
 \end{center} 
\end{figure}

Numerical results are presented in
Figures~\ref{2D_1:fig1},\ref{2D_2:fig2}. Fig.~\ref{2D_1:fig1} depicts
the time evolution of spatial diffusion coefficients for different
$p_{\parallel,0}/(m_ec)$ in a 2D approximate geometry.  Green curves
depict the time evolution in the parallel direction and blue curves
represent the same quantity in the transverse direction.
Displacements in the parallel direction are clearly ballistic:
$\langle \Delta x_{\parallel}^2 \rangle =\upsilon_{\parallel,0}^2
t^2$.  In the transverse plane, at least two regimes are identified:
for $t<t_{\rm nl}$, $\langle \Delta x_{\perp}^2\rangle \propto t^4$ as
expected from particle acceleration by nearly constant electric field
over its coherence cell, see Eq.~\ref{eq:dxt}; for $t>t_{\rm nl}$,
$\langle \Delta x_{\perp}^2 \rangle / \Delta t$ reaches a plateau
(i.e. diffusion) and its value has a power-law dependence on
$p_{\parallel,0}$ with slope of $-1/2$ (see the sub-panel of the
figure), which fits well the result of the previous Section, see
Eq.~\ref{eq:dt}. This behavior is a direct consequence of particle
trapping in every coherence cell it encounters, with a ``waiting
time'' per cell being equal to $t_{\rm nl}$. The case where
$p_{\perp,0}$ is different from zero was also investigated. In this
case there is no more trapping in the field coherence cell but random
motions in the transverse plane. We obtain a diffusive behavior on
longer time-scales as discussed in the previous Section (e.g
Eqs. \ref{eq:Dperp_OTSI_2 }, \ref{eq:Dperp_OTSI_3 }).

Figure~\ref{2D_2:fig2} presents the evolution of $\langle
p_{\parallel}^2\rangle/p_{\parallel,0}^2$ (green curves) and $\langle
p_{\perp}^2\rangle/p_{\parallel,0}^2$ (blue curves) as function of
time for different $p_{\parallel,0}/(m_ec) \in [1,10^4]$.  There is no
diffusion in momentum space with parallel component being bounded by
the $\pi_z$ invariant, while the transverse one varies as $\langle
\Delta p_{\perp}^2 \rangle /(m_ec)^2 \propto p_{\parallel,0}$. Since
the generalized momentum is constrained by this $\pi_{\rm z}$
invariant, we expect a transverse momentum gain $\langle \delta
p_{\perp}^2 \rangle = e E_{\perp}' \ell_{\perp} p_{\parallel,0}/c$.

Finally, we can mention that the energy gain $\langle \Delta
\gamma\rangle$ is independent of $p_{\parallel,0}$ and corresponds to
the amount of energy brought by the rms electric field potential. It
can be expressed as $\langle \Delta \gamma\rangle=e E'_{\perp}
\ell_{\perp}$. The energy variation from $\epsilon(p_{\parallel,0})$
to $\epsilon(p_{\parallel,0})+q E'\ell_{\perp}$ takes place when
$0<t<t_{\rm nl}$. For $t>t_{\rm nl}$, the particle energy remains
constant. 

\subsection{Transport in 3D-fields} \label{OTSI_3D_theory}

We consider now the particle transport over a longer time scale, for
the full 3D geometry, i.e. including a single resonant mode with $k_z
= \delta_e^{-1}$ in the case of OTSI and a continuum of small wave
numbers in the case of Weibel modes. In the wave frame, the
longitudinal coherence length $\ell_{\parallel}'\simeq\gamma_{\rm
  m}\ell_{\parallel}$, and $\ell_{\parallel}\sim \ell_{\perp}$ for
OTSI modes in the background plasma frame, so that
$\ell_{\parallel}'\simeq \gamma_{\rm m}\ell_{\perp}'$. For Weibel
modes, $\ell_{\parallel} \sim \delta_i/\xi_{cr}^{1/2}$ and
$\ell_{\perp} \sim \xi_B^{1/2} \delta_i$, thus
$\ell_{\parallel}/\ell_{\perp} \sim \xi_{cr}^{-1} $ and therefore
$\ell_{\parallel}'/\ell_{\perp}' \sim \gamma_{\rm m}/\xi_{cr} \gg 1$.

The invariance of the generalized z-momentum is now broken and the
particle momentum can diffuse in all directions. One limitation is the
phase-space confinement due to the total energy conservation, that
forbids energy diffusion. Another is that the longitudinal coherence
length is significantly larger than the coherence length in the
transverse direction. With respect to the above 2D analysis, we thus
expect a change of regime when the spatial variation in the $z$
direction is felt by the particle, which corresponds to time scales
much larger than $\ell_{\parallel}'/c$. The randomization of the
longitudinal component of the momentum is then expected over a time
$t_z$, with
\begin{equation}
t_z \simeq  {\ell_{\parallel}' \over c} \left( \frac{p_{\parallel,0} c}
  {eE_z' \ell_{\parallel}'}\right) ^2 \simeq \frac{\ell_{\parallel}'}{\ell_{\perp}'}  {\ell_{\perp}' \over c} \left(\frac{p_{\parallel,0} c}{\epsilon_\star}\right)^2 \ .\label{eq:tz}
\end{equation}
For both OTSI and Weibel modes, $E_z'\ell_{\parallel}'\simeq E_\perp'
\ell_{\perp}'$ because in the wave frame, $\bmath{\nabla}\times
\mathbf{E}'=0$. The above is a linear estimate of the scattering
timescale in the longitudinal direction, however simulations at low
energies suggest that the dependence on $\ell'_{\parallel}$ differs
from this latter. We will take into account that lack of knowledge by
introducing a factor $\chi$ measuring the delay, compared to the 2D
correlation time, of the full development of 3D-dynamics.  It turns
out to impact directly the maximal acceleration energy.

Over intermediate timescales, we expect to recover the previous
results about the transverse diffusion and no diffusion in the
longitudinal direction.  But on time scale longer than $t_z$, one
expects a 3D diffusion with
\begin{equation}
  D_{\perp} \,=\, {1\over 3} \ell_{\perp}' c \left({p_{\parallel} c
      \over \epsilon_\star}\right)^2\ ,\quad
  D_{\parallel} \,=\,\frac{\ell_{\parallel}'}{\ell_{\perp}'} D_{\perp} \label{eq:dtdl3d} \ .
\end{equation}
As before we assumed that the deflection over a coherence cell is
small with $\delta \theta \sim \epsilon_\star/ p_{\parallel,0}c$.  The
description of the expected behaviors of these diffusion regimes is
provided in the figures shown in the next paragraph.

\subsection{Numerical results in 3D-fields}
We take the same configuration as in the 2D case but with a finite
value of $E_{\parallel}'=E_{\perp}'/\gamma_{\rm m}$ in the case of
OTSI, with $k_\parallel'/k_\perp'=1/\gamma_{\rm m}$.  To simulate a
Weibel turbulence, we include an ensemble of modes, consistent with
the original wave equations of Weibel modes,
\begin{equation}
  E'_{\parallel}(k_{\parallel}') = -{k_{\parallel}' \over k_{\perp}^{'2}} (\bmath k_{\perp}' \times \bmath B'_{\perp}) 
  \cdot \bmath \beta_{\rm m} \ , 
\end{equation}
which insures the conservation of the total energy of each particle.
For this ensemble of Weibel modes, $k_{\parallel}'/k_{\perp}'$ takes
small values up to some parameter $\kappa<1$; consequently,
$E_\parallel'/E_\perp'\sim \kappa$.  Most numerical calculations were
done with the same time scales as in the 2D case.

\begin{figure}
\begin{center}
 \includegraphics[width=0.45\textwidth]{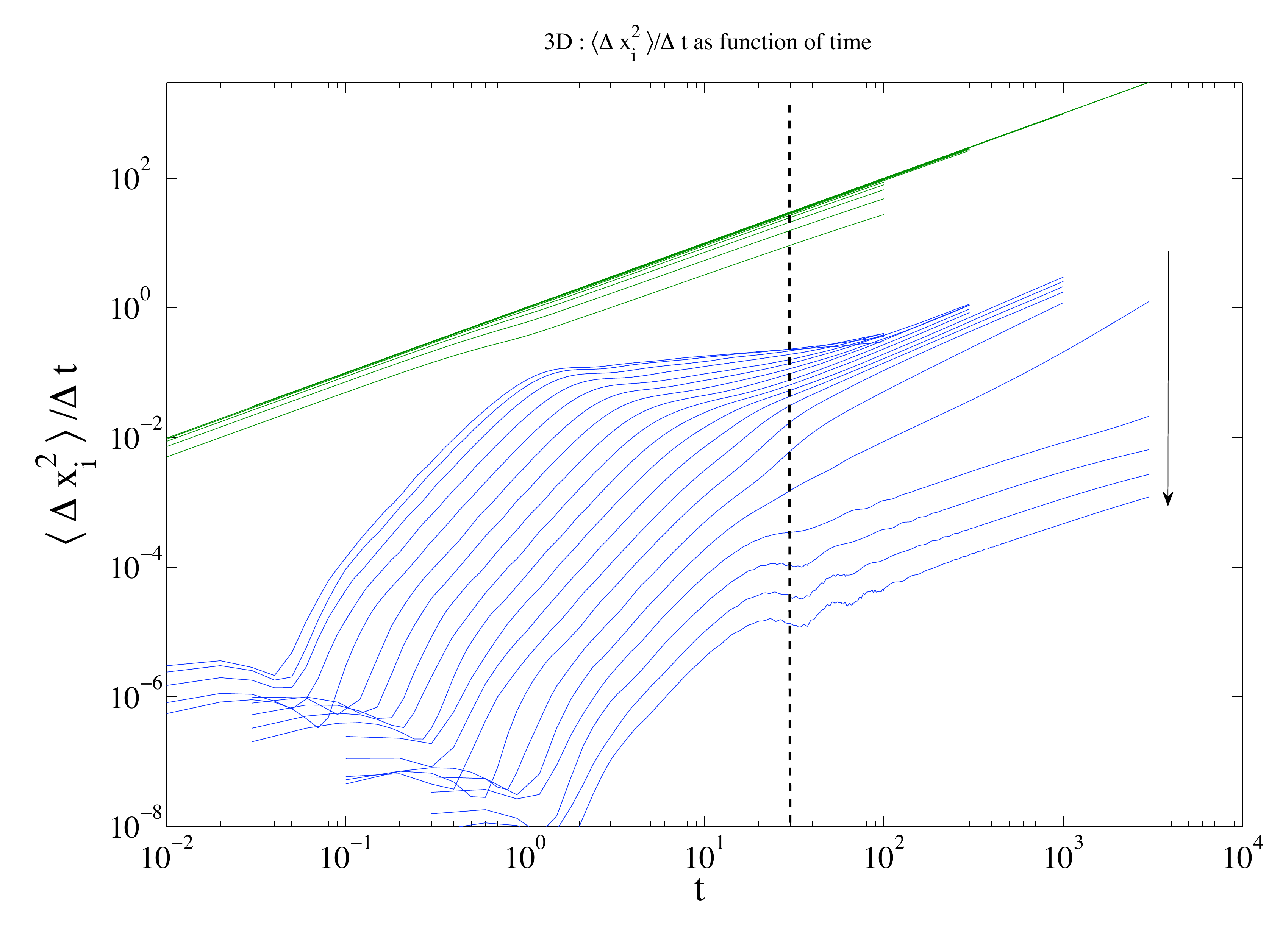}
 \caption{Transverse spatial diffusion in 3D modes of OTSI type.  This
   graph depicts the time evolution of spatial diffusion coefficients
   for different $p_{\parallel,0}/(m_ec)$ in 3D fields. Green curves:
   parallel ($z$) direction; Blue curves: transverse direction
   [($x,y$) plane]. Dashed vertical line delimits the linear coherence time
   in z direction $t_{\parallel,\rm c}=\gamma_{\rm m} l_{\perp}/c$.
    Displacements in the parallel direction remain
   ballistic on numerical time scales, since the integration time is
   too short to probe the diffusive behavior in this direction. In the
   transverse direction, 2D-like diffusive behavior (see
   Fig.~[\ref{2D_1:fig1}]) disappears gradually when $t_{\rm
     nl}<t<t_z$.}
  \label{3D_1:fig3}
 \end{center} 
\end{figure}

\begin{figure}
\begin{center}
 \includegraphics[width=0.45\textwidth]{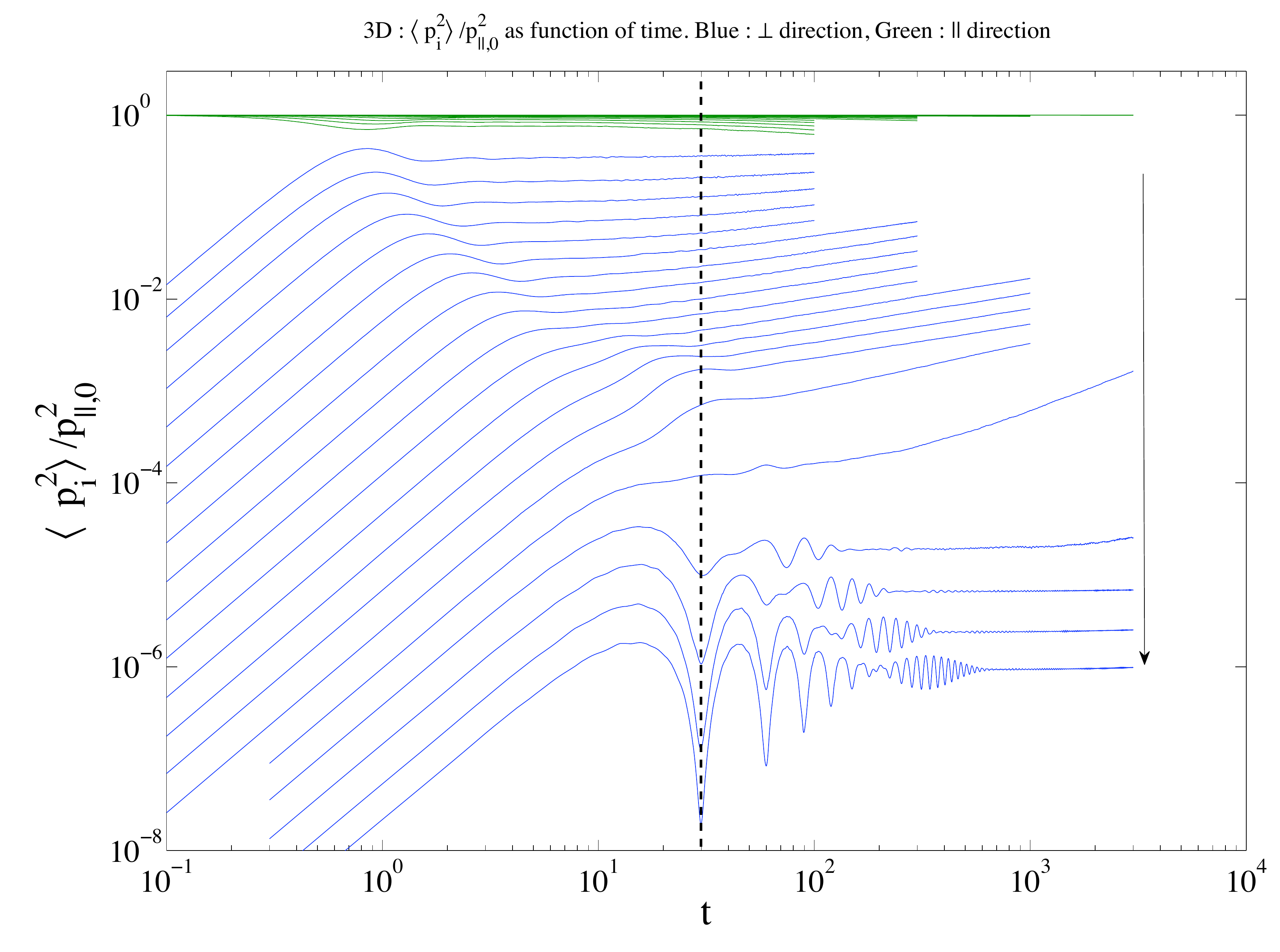}
 \caption{Time evolution of $\langle p_i^2 \rangle /p_{\parallel,0}^2$
   for different values of $p_{\parallel,0}/(m_ec)$, in 3D-fields.
   Green curves: parallel (z) direction; Blue curves: transverse
   direction. Dashed vertical line delimits the linear coherence time
   in z direction $t_{\parallel,\rm c}=\gamma_{\rm m} l_{\perp}/c$.
    Comparing to the Fig.~[\ref{2D_2:fig2}] $\langle
   p_{\parallel}^2 \rangle /p_{\parallel,0}^2$ keeps the same behavior
   as in 2D and the transverse components begin to rise at later times
   but longer time simulations are needed to explore its asymptotic
   behavior. A different behavior are observed for particles with
   momenta satisfying $t_{\rm nl}>t_{\parallel,\rm c}$, which are
   confined in a coherence cell in the transverse plane, see the text
   for details.}
  \label{3D_2:fig4}
 \end{center} 
\end{figure}

Figure~\ref{3D_1:fig3} depicts the time evolution of spatial diffusion
coefficients for different $p_{\parallel,0}/(m_ec)$ in 3D OTSI fields,
assuming $\gamma_{\rm m}=30$. Green curves correspond to the parallel
direction and blue curves to the transverse direction. Displacements
in the parallel direction remain ballistic on numerical time scales
and in the transverse direction the 2D-like diffusive behavior (see
Fig.~\ref{2D_1:fig1}) disappears gradually when $t_{\rm nl}<t<t_z$. It
will be seen further that the diffusive behavior is recovered in all
directions on longer time scales (i.e. $t\gg t_z$).
Figure~\ref{3D_2:fig4} presents the evolution of $\langle p_i^2
\rangle /p_{\parallel,0}^2$ in time for $p_{\parallel,0}/(m_ec) \in
[1,10^4]$.  Comparing to Fig.~\ref{2D_2:fig2}, one observes that
$\langle p_{\parallel}^2 \rangle /p_{\parallel,0}^2$ keeps the same
behavior as in 2D and the transverse components begin to rise at later
times; longer time simulations are needed to explore its asymptotic
behavior. Interestingly, a different behavior is observed for
particles with momenta satisfying $t_{\rm nl}>t_{\parallel,\rm c}$: in
this 3D OTSI turbulence, electromagnetic fields reverse every half
coherence length along the longitudinal direction, due to the
periodicity in that direction; therefore particles with $t_{\rm
  nl}>t_{\parallel,\rm c}$ execute oscillations in the transverse
direction, with ballisitic motion along the longitudinal
direction. Such particles are then confined in the transverse plane
because their motion is reversed before they have time to experience a
decorrelated field in the transverse plane. This is particularly
important with respect to acceleration efficiency, since such
particles would not return on a short timescale (in the absence of a
mean field, of course). One can check that the condition $t_{\rm
  nl}>t_{\parallel,\rm c}$ amounts to $p_{\parallel,0}> \epsilon_\star
\gamma_{\rm m}^2/2$ for $p_{\perp,0}=0$, which is easily satisfied, or
in terms of initial transverse velocity,
$p_{\perp,0}/p_{\parallel,0}<2/\gamma_{\rm m}^2$, which is also
generically satisfied for the suprathermal particles. This means that
OTSI turbulence is inefficient from the point of view of scattering
suprathermal particles away from the longitudinal direction.

\begin{figure}
\begin{center}
 \includegraphics[width=0.45\textwidth]{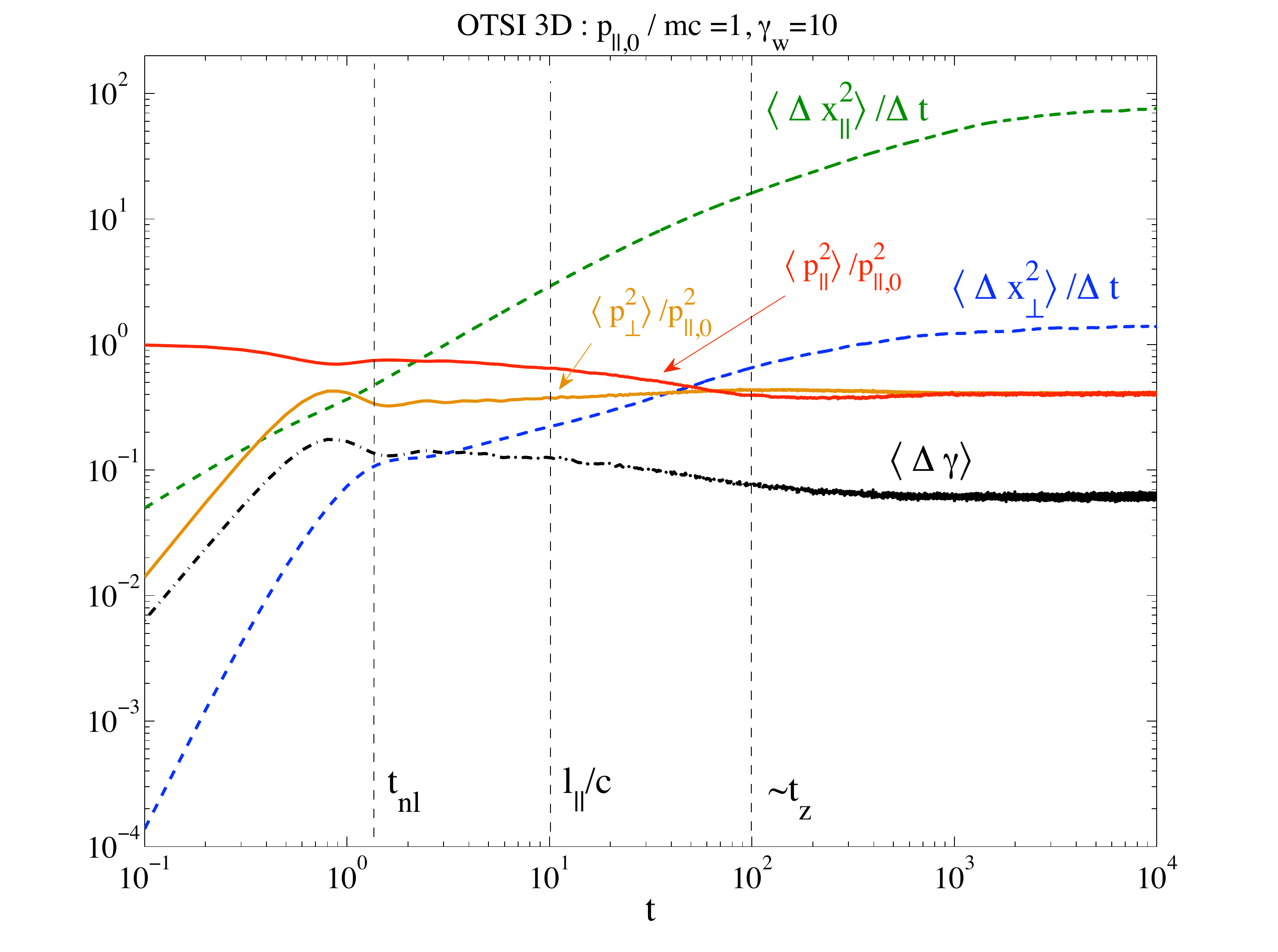}
 \caption{For OTSI modes, time evolution of different quantities in
   the case $p_{\parallel,0}=m_ec$ with enhanced integration time. Solid
   curves show the average $\langle p_i^2\rangle /p_{\parallel,0}^2 $,
   red color for parallel direction and orange color for the
   transverse one. Dashed curves show spatial transport coefficients
   $\langle\Delta x_i^2 \rangle / \Delta t$. Green color for parallel
   direction and blue for transverse direction. Dot-dashed black curve
   shows $\langle\Delta \gamma \rangle $. Vertical dashed lines indicate 
   tree caracteristic times relevant for particle dynamics. As expected, spatial
   diffusion is present in all directions on time scales much longer
   than $t_z$. Above the diffusion time all momenta reach isotropy:
   $\langle p_x^2 \rangle =\langle p_y^2 \rangle=\langle p_z^2 \rangle
   \simeq p_{\parallel,0}^2/3$.}
\label{3D_5:fig5}
 \end{center} 
\end{figure}

In order to test our estimates from subsection \ref{OTSI_3D_theory}, a
simulation with enhanced integration time was performed in the case
where $p_{\parallel,0}/m_ec=1$. The result is presented in
Fig.~\ref{3D_5:fig5} where different statistical quantities are
plotted as a function of time.  A slightly smaller value $\gamma_{\rm
  m}=10$ is adopted here to reduce the characteristic $t_z$ time. On
time scales larger than $t_z$ the spatial diffusion is recovered in
all directions, all momenta are isotropized and energy gain is equal
to the field rms energy. Note that the ratio between the spatial
diffusion coefficients in the parallel and transverse directions is
not equal to $\gamma_{\rm m}$ as expected in Eq.~\ref{eq:tz}: its
value is $55>\gamma_{\rm m}$ and scales as $0.6 \gamma_{\rm m}^2$ when
$\gamma_{\rm m}$ is varied explicitly in our simulations. It remains
uncertain if this scaling depends on the choice of low energy
particles [$p_{||,0}/(m_ec)=1$] and if it remains the same for highest
energies. Direct simulations for high energies, with large enough
integration time, are numerically prohibitive and are subject to
severe numerical errors. In the following, we encode this uncertainty
in a parameter $\chi$, which is a substitute for
$\ell_{\parallel}'/\ell_{\perp}'$, such that $\chi=\gamma_{\rm m}$ if
the linear value given by Eq.~\ref{eq:tz} were to apply, but $\chi\sim
\gamma_{\rm m}^2$ as indicated by the simulations.

\begin{figure}
\begin{center}
 \includegraphics[width=0.45\textwidth]{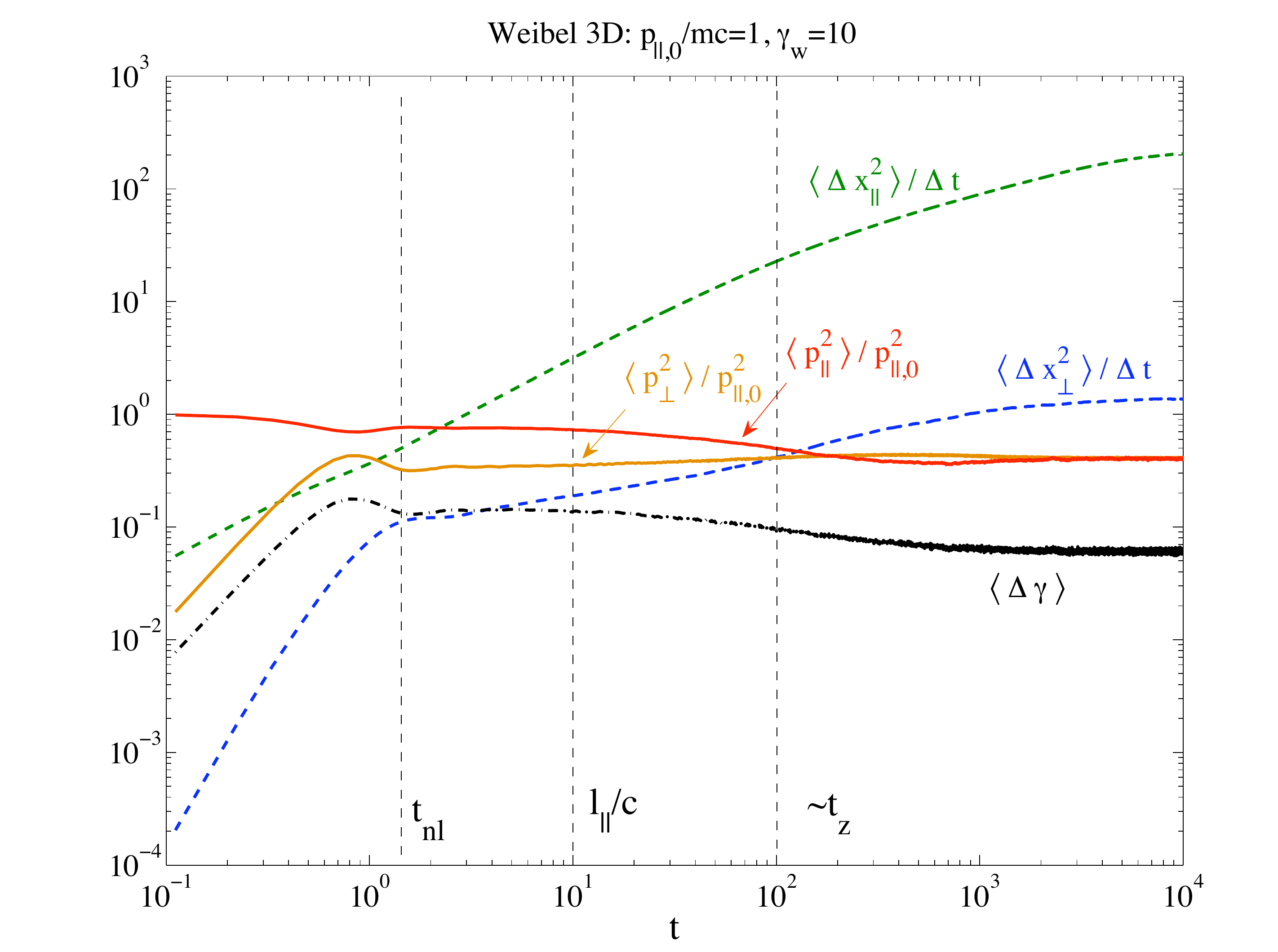}
 \caption{Same as in Fig.~\ref{3D_5:fig5}, but for Weibel 3D modes.
   Here, we take $k_{\parallel}'/k_{\perp}'=1/10$. This choice is
   dictated by numerical time limitation in order to observe the
   development of a 3D regime. As expected, spatial diffusion is also
   present in all directions on time scales much longer than
   $t_z$. Above the diffusion time all momentum components reach
   isotropy: $\langle p_x^2 \rangle =\langle p_y^2 \rangle = \langle
   p_z^2 \rangle \simeq p_{\parallel,0}^2/3$. We remark that the
   evolution is very similar to that obtained with OTSI modes, despite
   the finite range of $k_{z}'$ wavenumbers.  }
\label{3D_6:fig6}
 \end{center} 
\end{figure}

Finally, in the Fig.~\ref{3D_6:fig6} we present also the same
simulation but in the case of 3D Weibel-type fields with
$k_{\perp}'/k_{\parallel}'=10$, corresponding to $\kappa=0.1$.  The
general behavior is similar to that observed for transport in OTSI
turbulence. In particular, one recovers a scaling $t_z\propto
\kappa^{-2}$ and $D_\parallel/D_\perp\propto \kappa^{-2}$ instead of
the linear estimates given in Eqs.~\ref{eq:tz} and \ref{eq:dtdl3d},
which suggest a scaling in $\kappa^{-1}$. As before, we encode this
uncertainty with a factor $\chi$, so that $D_\parallel/D_\perp \simeq
\chi $, the simulations indicating $\chi \simeq
\left(\ell_\parallel'/\ell_\perp'\right)^2$.

As expected, the energy does not undergo a diffusive behavior. After a
limited gain, the particles keep a constant momentum.  This is in
agreement with theoretical predictions: stochastic acceleration is not
seen, transverse heating is bounded, the energetic particle beam is
broadened in the transverse direction, but the distribution remains
anisotropic.

\section{Electron heating} \label{sec4}

As we discussed in Section~\ref{sec2}, the wiggler parameter for the
background electrons in the proper frame of the microturbulent mode
is very large, actually $a\gg\gamma'$, with $\gamma'$ the typical
Lorentz factor of the electrons in that frame. Since the modes carry a
transverse electric field that is comparable to the magnetic field,
this offers a promising source of preheating in the shock precursor.
The growth of the electron temperature together with the growth of
electromagnetic waves gives rise to a DC-electric field in the normal
direction in order to maintain a stationary flow of electrons in the
front frame. In turn, this slows down the incoming proton flow and
electron preheating develops at the expense of the kinetic energy of
protons.

We first consider the effect of the electric field of the Weibel waves
in their proper frame.  The electron temperature temporarily
increases, as long as their energy is smaller than $e\Delta \Phi' = e
\bar E' \ell_{\perp}$, because, in this frame, the total energy of each
particle is conserved, as discussed before for suprathermal particles.
This limiting energy is
\begin{equation}
\label{ }
\bar \epsilon' =e\bar E'\ell_{\perp} = \xi_{\rm B}^{1/2} \gamma_{\rm m} {\ell_{\perp} \over \delta_{\rm i}} m_pc^2 \ .
\end{equation}
Reverting to the background plasma frame, that energy corresponds to a
temperature which is a sizable fraction of the proton energy
\begin{equation}
\label{TE1}
  T_{e {\rm,lim}} = \xi_{\rm B}^{1/2} \gamma_{\rm m} {\ell_{\perp} \over \delta_{\rm i}} m_pc^2\ ,\label{eq:te}
\end{equation}
assuming that most of the electron heating is distributed along the
transverse direction. The transverse characteristic size is inflated
by the high electron temperature, so that $\ell_{\perp} = \sqrt{\bar
  \gamma_e} \delta_e \simeq \sqrt{3T_e/m_pc^2} \delta_i$ (see
subsection 2.1 and appendix A).  Therefore the temperature in the
upstream co-moving frame is finally
\begin{equation}
\label{TE1}
  T_{e {\rm,lim \mid u}} = \xi_{\rm B} m_pc^2\ ,\label{eq:teu}
\end{equation}
and $\ell_{\perp} \sim \xi_B^{1/2} \delta_{\rm i}$.

This transitory heating process is not in a diffusion regime, it is
rather a direct linear acceleration in the coherent electric field of
a coherent cell. Using Eq.~\ref{eq:sys}, one can easily check that the
energy $\bar\epsilon'$ is obtained over a typical linear timescale
$\ell_{\perp}/c$. This fast heating in Weibel waves is a particular
case of a situation where thermal electrons undergo strong
relativistic motions in the waves, reaching relativistic temperatures
$T_e \sim a\, m_e c^2$. A similar process is at work in the OTSI
turbulence.

Because the transverse coherence scale is everywhere smaller than the
precursor length scale, on which $\xi_{\rm B}$ varies, the above fast
heating process brings forward a picture in which electrons are nearly
instantaneously heated to the local temperature given by
Eq.~\ref{eq:teu} above, scaling as $\xi_{\rm B}$. Of course, as the
electrons near equipartition with the ions, one may expect the above
heating process to saturate, notably because the oscillation parameter
then becomes smaller compared to $\gamma'$, as discussed in
Sec.~\ref{sec2}. Far from the shock, $\xi_{\rm B}\ll 1$ hence the
electrons are heated to temperatures well below $m_p/m_e$ (upstream
frame), while closer to the shock, the temperature
rises. Interestingly, PIC simulations indeed show a gradual evolution
of the electron temperature over the length scale of the precursor
(\citealp{SS11a}). The above preheating process provides a concrete
physical mechanism for this picture.

Note that our analysis differs from the recent proposal of
\cite{GSSB12}, which argues that the background electrons are heated
in the inductive longitudinal electric field of the filament. In the
present scenario, the electrons oscillate in the transverse electric
field in the wave frame, which corresponds through a Lorentz transform
to the transverse magnetic field of the filament in the upstream
frame.

The above estimate of the electron temperature allows to evaluate
several quantities of interest. The transverse equilibrium of the
filament may be described through the relation $\delta n T_e + \delta
B^2/8\pi = 0$, assuming that electrons share everywhere the same
temperature, with $\delta n$ the density contrast between the outside
and the inside of the filaments. This leads to $\xi_{\rm B}\,\approx\,
\left(\vert\delta n\vert/n\right)T_e/(m_p c^2)$, which, when compared
to the above estimate for $T_e$, suggests $\vert\delta n\vert/n\sim
1$. We also note that, using this estimate of the temperature, the
filament Lorentz factor becomes of order $\gamma_{\rm m} \sim
(\xi_{\rm cr} \xi_{\rm B})^{-1/4}$ close to equipartition and
$\xi_{\rm cr}^{-1/2}$ for $\xi_{\rm B} \leq \xi_{\rm cr}$, which
should thus increase with the distance to the shock. With $\xi_{\rm B}
\sim \xi_{\rm cr}$ at the shock front, as suggested in
Sec.~\ref{sec:stb} and as indicated by PIC simulations
(\citealp{SS11a}), one finds $\gamma_m \sim \xi_{\rm cr}^{-1/2}$ close
to the shock front.

The electron preheating has an important feedback on the instability:
although it does not saturate the instability, it reduces the phase
velocity of the Weibel modes, at least because it increases the small
mass ratio and determine the condition for the reflection of incoming
protons. Regarding OTSI modes, the heating process is similar but it
stops with the saturation of the OTSI when the electrons achieve a
relativistic temperature.

Across the shock front, the proton heating follows from the mixing of
the different proton flows carrying energy of order $\Gamma_{\rm s}
m_pc^2$ in the shock front frame. More puzzling is the issue of
electron heating. In the previous paragraph we saw that a significant
preheating of the electron is expected.  The final stage of electron
heating across the shock front is likely related to an effective Joule
effect due to scattering. The scattering frequency of thermal
particles in the shocked flow is $\nu_{\rm s*} \sim \xi_{\rm B}
\omega_{\rm pi}$ and the magnetic diffusivity $\nu_{\rm m} = \eta
c^2/(4\pi) = \nu_{\rm s*} \delta_{\rm i}^2 \sim \xi_{\rm B}
c\delta_{\rm i}$. The typical length of Joules dissipation is thus
$\ell_{\rm J} = 3\nu_{\rm m}/c \sim 3\xi_{\rm B} \delta_{\rm i}$,
which is a quite short distance for particle thermalization. A more
detailed estimate is obtained by looking at the absorption of each
Fourier mode, which leads to $\ell_{\rm J}(\lambda)= 3
\lambda^{2}/\xi_{\rm B} \delta_{\rm i}$. All the magnetic energy that
has been generated in the precursor is thus dissipated in the electron
heating, and the electron temperature remains in sub-equipartition
with that of protons, $T_e \sim \xi_{\rm B} \Gamma_{\rm s} m_pc^2$
(shock front frame).  It turns out that both processes, preheating and
Joules heating, have a similar contribution to the electron
temperature that corresponds to the dissipation of the magnetic energy
that has been generated by the Weibel instability.


\section{Application to acceleration at a gamma-ray burst external
  shock} \label{sec5}

Section~\ref{sec3} provides the tools required to discuss the
residence time of particles upstream of a relativistic shock. As
discussed in \cite{Aea01}, \cite{P9}, one must compare the diffusion
timescale in the microturbulence with the timescale associated to
rotation in the background field and keep the shorter of the two. Let
us discuss here the implications of the microturbulence. The Weibel
filaments are likely the best sites of scattering, since the OTSI
modes, at least in their linear description, lead to the confinement
of high energy particles along the shock normal.

The upstream residence time of suprathermal particles returning from
the shocked plasma into the upstream flow is that corresponding to a
deflection by an angle $1/ \Gamma_{\rm s}$ beyond which the shock
front catches up the particle (\citealp{GA99,Aea01}); this provides a
reasonable estimate of the acceleration time of the Fermi
process. This has to be estimated in the filament frame first, in
which the required deflection angle is $\sim1/\Gamma_{\rm f\mid m}$,
with $\Gamma_{\rm f\mid m}$ the relative Lorentz factor between the
shock front and the wave frame, as determined by Eq.~\ref{eq:gmf}. The
influence of a possible background magnetic field on the return
timescale is discussed further below. For a particle of energy
$\epsilon_{\rm m}$ in this wave frame, the residence time is
\begin{equation}
t_{\rm res,m} \sim \chi {\ell_{\perp} \over c} \left({\epsilon_{\rm
      \mid m} \over \epsilon_\star}\right)^2 
{1\over \Gamma_{\rm f\mid m}^2} \ ,
\end{equation}
where $\chi$ is a factor large compared to unity. It accounts for the
fact that the decorrelation time in the longitudinal direction is much
larger than that in the transverse direction, as discussed above. For
OTSI modes, the simulations indicate $\chi\sim \gamma_{\rm m}^2$,
although the confinement in the transverse plane leads to very
ineffective scattering; scattering is rather provided by the Weibel
modes, for which $\chi\sim \gamma_{\rm m}^2
k_\perp/k_\parallel>\gamma_{\rm m}^2$, in terms of the wavenumbers of
the instability measured in the background plasma rest frame.

Going to the front frame, one finds a residence timescale
\begin{equation}
  t_{\rm res,f} \sim \chi {\ell_{\perp} \over c} \left({\epsilon_{\mid m} \over \epsilon_\star}\right)^2 {1 \over \Gamma_{\rm f\mid m}^3} \ ,
\end{equation}
with $\epsilon_\star = (\gamma_{\rm m}/\Gamma_{\rm s}) e\bar B_{\rm
  \mid f} \ell_{\perp}$ in terms of the turbulent magnetic field in
the front frame $\bar B_{\rm \mid f}$, and $\epsilon_{\rm \mid f}
\simeq \Gamma_{\rm f\mid m}(1-\beta_{\rm f\mid m} ) \epsilon_{\rm m}$.
\begin{equation}
\label{TRES}
t_{\rm res \mid f} \sim  \chi {\ell_{\perp} \over c} 
\left( {\epsilon_{\rm \mid f} \over e\bar B_{\rm \mid f} \ell_{\perp}} \right)^2 {\Gamma_{\rm s} \over \gamma_{\rm m}} \ .
\end{equation}

The fact that the scattering takes place in a frame moving at high
speed shortens the residence time, but this gain is mitigated by the
anisotropy of the turbulent modes, which induces the $\chi$ factor.

\subsection{Electron acceleration at relativistic shocks and radiation}

We have developed arguments in favor of an efficient heating of the
electron fluid by the microturbulence, which confirms the idea that
the electrons could likely reach a sub-equipartition temperature at
relativistic shocks.  For instance at the external shock of a
gamma-ray burst (GRB), where the afterglow radiation is produced, the
electrons could achieve a temperature of a few tens of GeV. Indeed the
proton temperature is very high at the beginning of the afterglow, and
we have $ T_e \lesssim T_p \sim \Gamma_{\rm s} m_pc^2$, which
corresponds to a few tens of GeV. Intense short scale magnetic
turbulence develops because the interstellar magnetization parameter
is very low, $\sigma \sim 10^{-9}$.

What kind of radiation can be expected in such small scale field, much
more intense than the mean field? This depends on the wiggler
parameter $a$, now measured in the downstream frame:
\begin{equation}
\label{ }
a \equiv \frac{e\bar B \ell_{\rm c}}{m_ec^2} \sim \xi_{\rm B}^{1/2} \Gamma_{\rm s} {m_p \over m_e} \ .
\end{equation}
This parameter measures the capability of the magnetic force to
deviate a relativistic electron of Lorentz factor $\gamma$ by an angle
$1/\gamma$ (this is the reason for which $\gamma$ does not appear in
the definition). When $a >1$ the magnetic field produces a single
deviation of the electron in the emission cone of half angle $1/
\gamma$, whereas when $a<1$ the electron can undergo several wiggles
in the emission cone. When $a$ is large, the emission behaves like a
normal synchrotron radiation in a mean field, except that there is no
polarization. When $a$ is small, the emission is of jitter type
(\citealp{M0}). In the present case, the large wiggler parameter
ensures that the emission caused by shocked and accelerated electrons
at a relativistic shock is synchrotron-like; the analysis of the
emitted spectrum may provide a diagnosis of the magnetic turbulence
although the departures are expected to be moderate
(e.g. \citealp{KR10,FU10,Mea11}) and actually dominated by the decay
dynamics of the microturbulence downstream, which implies that
particles of different Lorentz factors cool in regions of different
magnetic field strengths (\citealp{L12}).

As for the suprathermal electrons, we find an estimate of the maximum
Lorentz factor, measured at shock front, achieved against synchrotron
loss; since the acceleration time $\propto \gamma^2/\bar B^2$ and the
synchrotron time $\propto \gamma /\bar B^2$, the maximum Lorentz
factor is independent of the magnetic field intensity. The estimate is
similar in spirit to that derived by \cite{KR10} up to the dependence
on $\gamma_{\rm m}$; note furthermore that these authors discussed the
downstream acceleration timescale, whereas we include the transport
into the upstream. The estimate in the front frame is:
\begin{eqnarray}
\label{eq:gmax}
\gamma_{\rm max} & \sim & \left(  \frac{4\pi e^2 \ell_{\rm c}}{\sigma_T m_e c^2}
{\gamma_{\rm m} \over \chi\Gamma_{\rm s}} \right)^{1/3} 
\sim   \left( \mu nr_e^3\right)^{-1/6} \left({\gamma_{\rm m} \over \chi\Gamma_{\rm s} } \right)^{1/3} \sim   \nonumber \\
& \sim & 7\times 10^6 \left( {\gamma_{\rm m} \over \chi\Gamma_{\rm s}}\right)^{1/3} \ .
\end{eqnarray} 
To obtain this result we have taken into account the electron
wandering upstream where it experiences a level of turbulence
comparable to the downstream one (as measured in the front frame);
this level of turbulence is comparable because the extension of the
electron trajectory upstream is much shorter than the high energy
proton trajectories that shape the precursor.  The corresponding
maximum energy of synchrotron photons is
\begin{eqnarray}
  \epsilon_{\gamma, \rm max} \sim \Gamma_s \gamma_{\rm max}^2 {\hbar e\bar B_{\mid f} \over m_e c^2} & \sim & \sqrt{\xi_{\rm B}} {\Gamma_{\rm s}^{4/3} (\gamma_{\rm m}/\chi)^{2/3} \over  
    (\mu nr_e^3)^{-1/6}} {m_p c^2 \over \alpha_f} \simeq \nonumber \\
  & \simeq & 3 \times \xi_{\rm B,-2}^{1/2}\Gamma_{\rm s,2.5}^2 n_0^{1/2}
  \left({\gamma_{\rm m} \over \chi\Gamma_{\rm s}}\right)^{2/3} \, {\rm GeV} \ ,
\end{eqnarray}
with the usual notation $\xi_{\rm B,-2}=\xi_{\rm B}/0.01$,
$\Gamma_{\rm s, 2.5}=\Gamma_{\rm s}/300$ and $n_0=n/1\,{\rm cm}^{-3}$,
and $\alpha_f \simeq 1/137$ the fine structure constant.  The maximum
photon energy appears stronger than that given by \cite{KR10}, because
the level of magnetic energy density in the external shock of a GRB is
proportional to the proton mass instead of the electron mass for which
only the MeV range would be reached.

The above estimate of $\gamma_{\rm max}$ balances the acceleration
timescale against the timescale for synchrotron losses in the
turbulent field. In principle, one should also include inverse Compton
losses, which puts strict constraints on the return timescale in the
upstream frame, as discussed by \cite{LW06}, \cite{LZ11}. This
requires to use the scattering timescale discussed above and follow
the proper treatment of Klein-Nishina suppression given in these
studies; this task is left for future work.

The above discussion considers an unmagnetized shock. In the presence
of a background magnetic field, return into the upstream can be achieved
through the rotation by an angle $1/\Gamma_{\rm s}$ in the background
field. The return timescale then corresponds to $t_{\rm res,0\mid
  f}\sim\epsilon_{\rm\mid f}/(\Gamma_{\rm s}e B_0)$ as measured in the
shock front frame, so that at the maximal Lorentz factor determined by
Eq.~\ref{eq:gmax}
\begin{equation}
  \frac{t_{\rm res,0\mid  f}}{t_{\rm res\mid f}}\sim
  2\left(\frac{\gamma_{\rm
        m}}{\chi}\right)^{2/3}\xi_{B,-2}B_{0,-5}^{-1}n_0^{2/3}\Gamma_{\rm
    s,2.5}^{1/3}\ ,
\end{equation}
with $B_{0,-5}=B_0/10\,\mu$G. This implies that the background
magnetic field starts to dominate the dynamics of the highest energy
electrons as soon as the ratio $\gamma_{\rm m}/\chi$ becomes
significantly smaller than unity, for the above fiducial values. Among
others, this guarantees that GeV photons can be produced,
independently of $\gamma_{\rm m}/\chi$.

Thus a single synchrotron-like spectrum extending up to several GeV,
even possibly a few tens, can be expected and thus is compatible with
observations. From that point of view, the efficiency of relativistic
shocks with respect to the production of high energy radiation can be
regarded as high.

\subsection{Relativistic shock and suprathermal protons in GRBs}

As the scattering time increases with $\epsilon^2$, the Fermi process
at relativistic shocks is not expected to be a fast accelerator at the
highest energies. For protons, acceleration is in general limited by
the dynamical time scale $r_{\rm s}/c$ of the shock in the laboratory
frame. For the external shock of a gamma-ray burst at the beginning of
the afterglow phase, the maximum energy achieved when the residence
time upstream balances the expansion time, is
\begin{eqnarray}
E_{\rm max} & = & 2 Z \Gamma_{\rm s} \left(\frac{\gamma_{\rm m}}{\chi}\right)^{1/2} \xi_{\rm B}^{1/2} 
\sqrt{r_{\rm s} \over  \delta_{\rm i}}\, m_pc^2  \nonumber \\
& \sim &  3.7 \times 10^{15}\times  Z \left( \frac{\gamma_{\rm
      m}}{\chi}\right)^{1/2} 
\Gamma_{\rm s,2.5}r_{\rm s,17} n_0^{1/4} \, \, eV \ .
\end{eqnarray}
Again, the above holds for an unmagnetized shock. As usual, $r_{\rm
  s,17}\,\equiv\,r_{\rm s}/10^{17}\,{\rm cm}$. The performance can be
improved if one takes into account a background magnetic field, which
leads to regular rotation and a shorter return timescale. The maximum
energy can then be written $E_{\rm max,0}=Z \Gamma_{\rm s} eB_0 r_{\rm
  s} \,\simeq\, 10^{16}\times Z\,B_{0,-6}r_{\rm s,17}\Gamma_{\rm
  s,2.5}\,$eV.  Thus although an energy of order $10^{16}$ eV is
achieved, the result is far from the range of so-called ultra-high
energy cosmic rays, see also \cite{GA99}.

\section{Conclusions and prospects}
The development of a collisionless shock involves three essential
interrelated ingredients: the generation of suprathermal particles,
the generation of magnetic turbulence, the building up of a reflecting
barrier for a part of the incoming particles. This paradigm applies
successfully to non-relativistic as well as to relativistic weakly
magnetized shock waves.  Numerical and theoretical works have made
significant progress in understanding the physics and in providing
quantitative results that become useful for astrophysical
investigations. This includes not only the spectrum index and cut off
of the distribution of accelerated particles, but also the conversion
factors into cosmic rays, magnetic turbulence and radiation.

In this paper we have presented new theoretical investigations
regarding the transport of suprathermal particles in the
microturbulence upstream of the relativistic shock, and the preheating
of the background electrons. We have placed emphasis on the fact that
the microturbulent modes actually move relatively the background
plasma, with a possibly large Lorentz factor depending on the
background electron temperature. This motion of the microturbulence
generates a motional electric field in the frame in which the
filaments are static, which leads to fast heating of the background
electrons through relativistic oscillations.  Despite that the Weibel
instability generates magnetic filaments -- in the background plasma
frame -- whereas the oblique two stream instability generates almost
electrostatic waves, they behave similarly in their proper frame, in
which they are composed of an electrostatic field and a magnetostatic
field of almost the same amplitude.  This heating mechanism is
particularly efficient: within a transverse coherence length of the
perturbations, it heats the electrons to $\sim \xi_{\rm B} m_pc^2$, in
which $\xi_{\rm B}$ should be understood as the local (position
dependent) fraction of energy density stored in the electromagnetic
component. Because the coherence length is much shorter than the size
of the precursor, this brings forward the picture in which the
electrons are instantaneously heated to the above temperature, so that
their temperature rises gradually towards near equipartition as they
approach the shock front, a picture which appears in satisfactory
agreement with the results of \cite{SS09}, \cite{SS11a}. As we have
discussed, one should expect $\xi_B\sim\xi_{\rm cr}$ at the shock
front, from the condition that the Weibel turbulence has become
sufficiently strong to reflect the incoming particles. The Weibel
turbulence thus apparently draws the maximum amount from the
suprathermal particle energy reservoir, in qualitative agreement with
PIC simulations.

Electron preheating modifies the generation of microturbulence: it
saturates the oblique two stream instability and slows down the
propagation of Weibel modes. So we envisage that the nose of the
precursor contains fast propagating Weibel modes and then, closer to
the shock front, relativistic thermal electrons that enlarge the
characteristic scale.  The oblique two stream remain however likely
active in the cold phase at the tip of the shock precursor, like
Buneman instabilities which also preheat the electrons.

We have also discussed in some detail the properties of transport of
the suprathermal particles in the microturbulence. The filamentary
nature of the magnetic filaments strongly limits the scattering of
these particles in the longitudinal direction. The acceleration
process is accordingly slowed down by the time it takes for the
particle to probe effectively the inhomogeneities in the longitudinal
direction, as quantified here by the factor $\chi$. This strongly
suggests that PIC simulations of the Fermi process in 2D probably
involves mirror effects on the shock front rather than actual
upstream/downstream scattering, especially at the ``low'' energies
corresponding to the first Fermi cycles probed by these
simulations. To probe the 3D scattering regime discussed here, one
would need 3D PIC simulations with very long integration timescales,
in order to accelerate particles to energies such that their Larmor
radius in the turbulent field becomes larger than the coherence
length.

Shocks in AGN, blazar jets, or in the internal flow of GRBs are mildly
relativistic and therefore not subject to the severe restriction
imposed to the Fermi process by the mean field as it happens in the
ultra-relativistic regime. Thus, as argued here and in,
e.g. \cite{L9b}, \cite{PL11b}, those objects are better
candidates as sources of ultra-high energy cosmic rays.  In pulsar
wind nebulae, reconnections likely contribute to injecting high energy
particles in the shock and a suprathermal tail with a hard component
may be generated (\citealp{L03,PL07,SS11b}). At the weakly magnetized
external shock of a gamma-ray burst, Fermi acceleration should be
operative and then lead to extended synchrotron spectrum up to GeV
energies; although, if the shock propagates in a sufficiently
magnetized circumburst environment, the Fermi process may be, in a
first step, quenched by the mean field, which would lead to distinct
signatures (\citealp{LP11b}).

\section*{Acknowledgement}
It is a pleasure to acknowledge fruitful discussions with Lorenzo
Sironi and Brian Reville, supported by ISSI.  This work has been
financially supported by the GdR PCHE and the GDRE "Exploring the Dawn
of the Universe with Gamma-Ray Bursts" and by the PEPS-PTI Programme of
the INP/CNRS.


\appendix

\section{Weibel instability with a non-vanishing parallel wavenumber}

Despite detailed analyses of the relativistic beam instability in the
Weibel regime, e.g. \cite{WA04}, \cite{AW07a}, \cite{AW07b},
\cite{Sea11} and \cite{BFP05}, the motion of filaments has not been
given attention so far. We will emphasize this issue for a cold and a
relativistically hot electron fluid.

For a cold background plasma pervaded by a cold monokinetic beam, the
wave system is described by the matrix:
\begin{equation}
\Lambda_{ij} = \left( 1-{\omega_{\rm p}^2 \over \omega^2} \right)-{k^2c^2 \over \omega^2} \left(\delta_{ij}-{k_ik_j \over k^2}\right) + \chi^{\rm b}_{ij} \ ,
\end{equation}
where $\chi^{\rm b}_{ij}$ are the components of the susceptibility
tensor of the beam plasma given by
\begin{equation}
\chi^{\rm b}_{ij} = -\xi_{\rm cr} \mu {\omega_{{\rm p}e}^2 \over \omega^2} \left( \delta_{ij}+ \frac{k_i \upsilon_j+k_j \upsilon_i}{\omega-\bmath k \cdot \bmath \upsilon_{\rm b}} + \frac{k^2c^2-\omega^2}{(\omega-\bmath k \cdot \bmath \upsilon_{\rm b})^2}{\upsilon_i\upsilon_j \over c^2} \right) \ .
\end{equation}
We chose a beam velocity direction along $\mathbf{z}$, $\bmath
\upsilon_{\rm b} = \upsilon_{\rm b} \bmath e_z$ and a wave vector
$\bmath k =k_x \bmath e_x + k_z \bmath e_z$.  Electromagnetic waves
polarized in the y-direction are decoupled. The wave system of
interest reduces to second order, such that
\begin{eqnarray}
\Lambda_{xx} & = & 1-{\omega_{{\rm p}e}^2 \over \omega^2} \left( 1+k_z^2 \delta_e^2+\xi_{\rm cr} \mu \right) \nonumber \\
\Lambda_{xz} = \Lambda_{zx} & = & {k_xk_zc^2 \over \omega^2} - \xi_{\rm cr} \mu {\omega_{{\rm p}e}^2 \over \omega^2}{k_x\upsilon_{\rm b} \over \omega-k_z\upsilon_{\rm b}} \nonumber \\
\Lambda_{zz} & = & 1-{\omega_{{\rm p}e}^2 \over \omega^2} \left( 1+k_x^2 \delta_e^2 + \xi_{\rm cr}\mu \right) \nonumber \\
  & & - \, \xi_{\rm cr} \mu {\omega_{{\rm p}e}^2 \over \omega^2}
  \left( \frac{2k_z\upsilon_{\rm b}}{\omega-k_z\upsilon_{\rm
        b}}+\frac{k^2c^2-\omega^2}{(\omega-k_z\upsilon_{\rm b})^2}
    \beta_{\rm b}^2 \right) \ .
\end{eqnarray}
The dispersion relation reads
\begin{equation}
\label{ }
{\cal D}(k, \omega)= \Lambda_{xx} \Lambda_{zz} - \Lambda_{xz}^2 = 0\ .
\end{equation}

In the case of Weibel instability, $\omega^2 \ll \omega_{{\rm p}e}^2$,
$k_x^2 \delta_e^2 \sim 1$ and $k_z^2 \delta_e^2 \ll 1$, we find:
\begin{equation}
\label{ }
\omega = k_z \upsilon_{\rm b} \left( 1-\xi_{\rm cr} \mu\frac{\gamma_{\rm b}^{-2}+k_{\perp}^2\delta_e^2}{1+k_{\perp}^2\delta_e^2} \right) + i \sqrt{\xi_{\rm cr} \mu} \frac{k_{\perp}\upsilon_{\rm b}}{(1+k_{\perp}^2 \delta_e^2)^{1/2}} \ ,
\end{equation}
with $k_{\perp}=k_x$.  This is the result for a cold background plasma
and a cold beam.  When one takes into account a dispersion of the beam
within an angle $1/\Gamma_{\rm s}$, as previously investigated
(\citealp{Rea11,LP11a}), the instability is quenched when $\Gamma_{\rm
  s} < ({\xi_{\rm cr} \mu})^{-1/2}$. However the instability is
restored when the background electrons are sufficiently hot with both
the same growth rate and the same frequency.  In the case of OTSI,
electron heating up to relativistic temperature tends to quench the
instability because the modes become superluminal and thus the
resonant interaction cannot be achieved; this effect is however
delayed by the fact that the frequency is negatively shifted by the
beam, which lowers the phase velocity.

Let us now study a problem similar to the above, albeit for a
relativistic electron temperature. As the electrons can come close to
equipartition, it is essential to account for the response of the
background ions. We thus write the components of the wave tensor:
\begin{eqnarray}
  \Lambda_{xx} & = & \varepsilon_{\parallel} {k_x^2 \over k^2} + \left( \varepsilon_{\perp} -\eta^2\right) {k_z^2 \over k^2} + \chi^{\rm b}_{xx} \nonumber \\
  \Lambda_{xz} = \Lambda_{zx} & =  & \left( \varepsilon_{\parallel} -\varepsilon_{\perp} + \eta^2\right) {k_xk_z \over k^2} + \chi^{\rm b}_{xz} \nonumber \\
  \Lambda_{zz} & = & \varepsilon_{\parallel}{k_z^2 \over k^2} + \left( \varepsilon_{\perp}-\eta^2 \right) {k_x^2 \over k^2} + \chi^{\rm b}_{zz} 
\end{eqnarray}
where the dielectric coefficients for relativistically hot electrons
and cold protons in the low frequency approximation are:
\begin{eqnarray}
  \varepsilon_{\parallel} & \simeq & 1-{\omega_{\rm pi}^2 \over
    \omega^2} + {1\over k^2 \lambda_{\rm De}^2} \left( 1+i{\pi \over 2}{\omega \over kc}\right) \\
  \varepsilon_{\perp} & \simeq & 1-{\omega_{\rm pi}^2 \over \omega^2} -
  {1\over k^2 \lambda_{\rm De}^2} \left( 1 -i{\pi \over 4}{kc \over
      \omega}\right) \ .
\end{eqnarray}
When considering relativistically hot electrons, it is convenient to
write the Debye length $\lambda_{\rm De}$ such that
$$
\lambda_{\rm De}^2 \equiv {T_e \over 4\pi ne^2} = \bar \mu \delta_{\rm
  i}^2 \, \, {\rm with} \, \, \bar \mu \equiv \frac{1}{3}\bar \gamma_e\mu \, \,
{\rm and} \, \, \bar \gamma_e \equiv 1+3T_e/m_ec^2 \ .
$$
Whereas the Landau contribution (imaginary part) is a small correction
to the longitudinal response, it is dominant in the transverse
response.
\begin{eqnarray}
  \omega^2\Lambda_{xx} & = & \omega^2 \left( 1+\frac{\omega_{\rm pi}^2}{\bar \mu k^2c^2} \right)-\omega_{\rm pi}^2 \left( 1+\xi_{\rm cr} \right) \nonumber \\
  \omega^2\Lambda_{zx} & = & {\omega^2 \over k^2 \lambda_{De}^2}\left( 2-i{\pi \over 4}{kc \over \omega} \right){k_xk_z \over k^2}+ k_xk_zc^2 
  - \, \xi_{\rm cr} \omega_{\rm pi}^2 {k_x\upsilon_{\rm b} \over \omega-k_z\upsilon_{\rm b}} \nonumber \\
  \omega^2\Lambda_{zz} & = & \omega^2 \left( 1-{\omega_{\rm pi}^2 \over \bar \mu k^2c^2} \right)+i{\pi \over 4} {\omega_{\rm pi}^2 \over \bar \mu}
  {\omega \over kc}  \nonumber \\
  & & - \, \omega_{\rm pi}^2 \Bigg( 1+k_x^2 \delta_{\rm i}^2 + \xi_{\rm cr} + \xi_{\rm cr}\frac{2k_z\upsilon_{\rm b}}{\omega-k_z\upsilon_{\rm b}} 
  + \xi_{\rm cr}\frac{k^2c^2-\omega^2}{ \left(
      \omega-k_z\upsilon_{\rm b}\right)^2} \beta_{\rm b}^2 \Bigg) \ ,\nonumber\\
 &&
\end{eqnarray}
with $\Lambda_{xz}=\Lambda_{zx}$.  An important observation is that
$\omega_{\rm pb} < \omega_{\rm pi}$, so that one can neglect
$\vert\omega\vert^2$ in front $\omega_{\rm pi}^2$. Furthermore, at
equipartition, the ions contribute strongly to the instability and the
typical wavenumber $k_x\sim \omega_{\rm pi}$, whereas for
$\bar\mu\ll1$, the response of the electrons dominate, and
$k_x\rightarrow \bar\mu^{-1/2}\omega_{\rm pi}$, the latter
corresponding to the relativistic electron plasma frequency. Thus the
dispersion relation can be written (omitting also a term in $\xi_{\rm
  cr}^2$, and assuming $k_z^2 \delta_{\rm i}^2 \ll 1$):
\begin{eqnarray}
\label{ }
\left(1-\frac{\omega^2}{\bar \mu k^2c^2}\right) \left[ -i{\pi \over 4}{\omega \over \bar \mu kc} +1+k^2\delta_{\rm i}^2 + 
  \xi_{\rm cr} \left({k^2c^2 \over \delta \omega^2} + {2\over \gamma_{\rm b}^2}{k_z\upsilon_{\rm b} \over \delta \omega}\right) \right] \nonumber\\
+ 2\xi_{\rm cr} {k_z\upsilon_{\rm b} \over \delta \omega} \left( k_x^2 \delta_{\rm i}^2-i{\pi \over 4}{\omega \over \bar \mu kc}{k_x^2 \over k^2} \right) = 0
\end{eqnarray}
We solve that equation by setting $\delta \omega = \nu + i \gamma$
with the approximation $\vert \nu \vert \ll \vert\gamma\vert$
($\gamma$ should not be confused with Lorentz factors appearing
elsewhere).  The growth rate is the positive root of the equation
\begin{equation}
  \label{ }
  {\pi \over 4} \frac{\gamma^3}{\bar \mu k^3c^3} + \left( 1+k^2\delta_{\rm i}^2 \right) \frac{\gamma^2}{k^2c^2} - \xi_{\rm cr} = 0 \ ,
\end{equation}
which gives
\begin{equation}
  \gamma\,\simeq\, \sqrt{\xi_{\rm cr}}\upsilon_{\rm b}\omega_{\rm
    pi}\,\frac{k_x\delta_i}{\left[1+(k_x\delta_i)^2\right]^{1/2}}\ ,
\end{equation}
provided $(k_x\delta_i)^3\gtrsim \sqrt{\xi_{\rm cr}}/\bar\mu$. If
$\bar\mu>\sqrt{\xi_{\rm cr}}$, the latter inequality is verified for
$k_x\delta _i\gtrsim 1$, which means the filamentation instability is
of the ion-ion type, with typical wavenumber $k_x\sim \delta_i^{-1}$.
This represents the range of wavenumbers for which the growth rate
peaks, because if $(k_x\delta_i)^3 < \sqrt{\xi_{\rm cr}}/\bar\mu$ the
Landau effect on hot electrons reduces the growth rate to $\gamma
\simeq (4/\pi \bar \mu \xi_b)^{1/3}kc$.  Now, if
$\bar\mu<\sqrt{\xi_{\rm cr}}$, the instability is pushed towards
higher values of $k_x$, as the response of the hot electrons
dominates.

The  frequency shift is obtained at the first order of the expansion in
$\nu$:
\begin{equation}
\label{ }
\nu\,\simeq\, -k_z\upsilon_{\rm b} \displaystyle{\frac{\frac{\pi}{8}\gamma +
    (k_x\delta_i)^3\bar\mu\xi_{\rm cr}\omega_{\rm pi}}
{\frac{3\pi}{8} \gamma +
  (k_x\delta_i)\left[1+(k_x\delta_i)^2\right]\bar\mu\omega_{\rm pi}}}\ .
\end{equation}
Using the value of $\gamma$ and simplifying $1+(k_x\delta_i)^2\sim
(k_x\delta_i)^2$, one obtains
\begin{equation}
\nu\,\simeq\, -k_z\upsilon_{\rm b}\sqrt{\xi_{\rm cr}}\frac{\frac{\pi}{8} +
  (k_x\delta_i)^3\bar\mu\sqrt{\xi_{\rm cr}}}{\frac{3\pi}{8}\sqrt{\xi_{\rm cr}} +
    (k_x\delta_i)^3\bar\mu}\ ,
\end{equation}
which takes different scalings, depending on the comparison between
$\bar\mu$ and $\xi_{\rm cr}$.

If $\bar\mu>\sqrt{\xi_{\rm cr}}$, the filamentation instability grows
at $k_x\delta _i\sim 1$, so that
\begin{equation}
\nu\,\simeq\, -k_z\upsilon_{\rm b}\frac{\pi}{8}\frac{\sqrt{\xi_{\rm
      cr}}}{\bar\mu}\quad\quad\left(\bar\mu>\sqrt{\xi_{\rm cr}},\,\,
  k_x\delta_i\sim1\right)\ .
\end{equation}

If $\bar\mu<\sqrt{\xi_{\rm cr}}$, maximum growth takes place at
$k_x\delta_i\simeq \bar\mu^{-1/2}$, so that
\begin{equation}
  \nu\,\simeq\, -k_z\upsilon_{\rm
    b}\left(\frac{\pi}{8}\sqrt{\bar\mu\xi_{\rm cr}}+\xi_{\rm cr}\right)
\quad\quad\left(\bar\mu<\sqrt{\xi_{\rm cr}},\,\,
    k_x\delta_i\sim \bar\mu^{-1/2}\right)\ ,\label{eq:nulowmu}
\end{equation}
although it should be noted that the growth rate increases weakly with
$k_x\delta_i$, and that the frequency shift evolves in a non-trivial
way with $k_x$ in the interval $\xi_{\rm
  cr}^{1/6}\bar\mu^{-1/3}\omega_{\rm pi} \rightarrow
\bar\mu^{-1/2}\omega_{\rm pi}$.

We expect that the size of the Weibel filaments, i.e. the transverse
coherence length is determined by the maximum $k_x$; thus
$\ell_{\perp} \sim \delta_i \sqrt{\bar \mu}$ far from equipartition,
with $\ell_{\perp}\rightarrow\delta_i$ close to equipartition. For
this typical size, the filament Lorentz factor goes from $\gamma_{\rm
  m}\sim \xi_{\rm cr}^{-1/2}$ when $\bar\mu<\xi_{\rm cr}$ to
$\gamma_{\rm m} \sim (\bar \mu \xi_{\rm cr})^{-1/4}$ when $\xi_{\rm
  cr}\lesssim\bar\mu\lesssim \sqrt{\xi_{\rm cr}}$, and to $\gamma_{\rm
  m}\sim \xi_{\rm cr}^{-1/4}\bar\mu^{1/2}$ for $\bar\mu>\sqrt{\xi_{\rm
    cr}}$.

The wavelength in the normal direction is limited by $k_z
\upsilon_{\rm b} < \gamma$ which implies $\ell_{\parallel} >
\delta_{\rm i}/\sqrt{\xi_{\rm cr}}$, which is comparable with the
growth length.

\end{document}